\documentclass[%
 reprint,
 amsmath,amssymb,
 aps,
]{revtex4-2}

\usepackage{graphicx}
\usepackage{dcolumn}
\usepackage{bm}

\begin{document}


\title{On void formation during the simulated tensile testing of polymer-filler particle composites}

\author{John J. Karnes}
 \email{karnes@llnl.gov}
\author{Supun S. Mohottalalage}%
\author{Amitesh Maiti}
\author{Andrew P. Saab}
\author{Todd H. Weisgraber}

\affiliation{%
 Lawrence Livermore National Laboratory\\
 5000 East Ave., Livermore, CA, 94550, USA
}%




\date{\today}

\begin{abstract}
We simulate a series of model polymer composites, composed of linear polymer strands and spherical, monodisperse filler particles (FP). These molecular dynamics simulations implement a coarse-grained, bead-spring force field and we vary several formulation parameters to study their respective influences on material properties. These parameters include FP radius, FP volume fraction, temperature, and the polymer-polymer, FP-FP, and polymer-FP interaction potentials. Uniaxial extension of the simulation cells allows direct comparison of mechanical reinforcement (or weakening) provided by the FP. We focus on the formation of microscopic voids during simulated tensile testing of glassy polymer composites and quantify how the characteristic spatial and morphological arrangement of these voids is a function of interaction potentials used in the simulation. We discuss the implications of these findings in the context of polymer composite design and formulation.

\end{abstract}

\maketitle

\section{\label{sec:intro}Introduction}

Polymers are a unique class of materials, defined by the chains of covalently-bonded, repeating sub-units that comprise them~\cite{staudinger_uber_1920, mulhaupt_hermann_2004}. Now commonplace, the `macromolecular hypothesis' of the material's microscopic structure met strong resistance when first proposed by Staudinger in 1920~\cite{rubinstein_polymer_2003}. The chemical makeup of these sub-units, length of the chains, and crosslinking and entanglement of the chains are a few of the tunable parameters that affect the properties of the resulting material. This parameter space allows for the design of polymers optimized for a wide range of properties, including tensile strength, thermal conductivity, electrical insulation, resistance to chemical degradation, or wetting. Further, polymer morphology can be tuned for more esoteric properties based on structure, including the synthesis of nanoporous materials with low areal densities that permit them to function as surrogates for high density gases~\cite{tillotson_transparent_1992,pekala_carbon_1992, pekala_structure_1993}. The ever expanding set of macroscopic properties that emerge from the interplay of microscopic interactions continues to inspire interest in polymers whenever a bespoke set of properties is needed for a new application.

Beyond the design and formulation of the polymer, the addition of distinct filler materials like carbon fibers or silica particles to the polymer matrix can be exploited to tune material properties outside the range achieved by reformulation of the polymer alone~\cite{witten_reinforcement_1993,yadav_role_2023,jouault_WellDispersed_2009,maiti_mullins_2014,rueda_rheology_2017}. The filler material's geometry, chemistry, and interaction with the surrounding polymer matrix all affect the properties of the resulting composite, and these changes may be desirable or detrimental as compared to the neat polymer for a given application~\cite{domurath_concept_2017}. In many applications nanoparticles are incorporated into a polymer matrix to enhance the mechanical performance of the material~\cite{witten_reinforcement_1993,fu_effects_2008}. This approach has proven effective for many use cases, despite the ongoing debate regarding the origin of the enhancement~\cite{vacatello_molecular_2002,lin_origin_2020,sun_molecular_2021,shi_molecular_2023,shen_revisiting_2020}. 

Lattice Monte Carlo studies have suggested that spherical filler particles (FP) with radii on the order of the radius of gyration $R_g$, provide mechanical reinforcement by facilitating the creation of transient polymer networks that bridge the FP, that this reinforcement is a strong function of the mean interparticle distance~\cite{ozmusul_lattice_2005}. Later experimental studies have proposed that mechanical reinforcement in composites is due to the formation of FP agglomerates and that bulk deformation causes the breakdown and subsequent reformation of these agglomerates~\cite{shui_intrinsic_2021}. It has been suggested that the force driving this agglomeration increases linearly with FP radius, independent of any explicit interparticle attractive potential~\cite{moczo_particulate_2016}. More recent molecular dynamics simulations performed in the glassy state agree with earlier computational work, and suggest that mechanical enhancement by FP arise from the polymer bridging effect, even in the case of polymers well below the entanglement length~\cite{lin_origin_2020, lin_chain_2021}. Deep understanding of these microscopic phenomena can assist the development of functional composite materials, particularly when balancing physical and economic drivers.  

The vastness of polymer design space is a challenge onto itself due to the \textit{formulate-fabricate-test} nature of experimental work. Due to practical limitations of chemical synthesis and formulation, thorough experimental parameter sweeps are both resource-intensive and often incomplete. Computer simulation is an alternative approach toward surveying this parameter space that can circumvent some of the challenges of an experimental campaign. All-atom, particle-based molecular dynamics simulations have demonstrated the ability to capture important properties of polymer melts and materials~\cite{afzal_HighThroughput_2021, bowman_free_2019,estridge_effects_2018,karnes_network_2020} but similar studies of polymer composites require access to larger time and length scales, making all-atom simulations of these systems inaccessible with modern computational resources~\cite{voth_CoarseGraining_2008,pavlov_fully_2016,alessandri_martini_2021}. Successful simulation of polymer composites must balance efforts to accelerate calculations while retaining the fidelity required by the specific problem. Coarse grained models reduce the cost of computer simulation by merging multiple atoms into a single interaction site. This approach reduces computational expense in two fundamental ways: First, reducing the number of particles in the system directly reduces the number of pairwise interactions that must be computed. Secondly, coarse-graining typically removes the highest frequency vibrations in a simulation. This allows the simulation to be performed with a longer time step since the maximum length of the time step is dictated by the fastest motion in the system.

Exploring polymer design space with computer simulation enables a major philosophical shift, where fundamental parameters like polymer chain length or interaction energies may be adjusted without having to first consider wet chemical synthesis or formulation; these challenges can be reserved for when promising candidate systems are identified. Similarly, coarse-grained simulations that implement well-validated models can provide molecular insight into the origins of the macroscopic properties that they model~\cite{karnes_isolating_2023, lin_origin_2020}.

In this work, we present a series of simulations that surveys parameters of interest to the polymer composite community: Filler particle (FP) size, FP volume fraction, polymer-FP interaction energy, and system temperature. We implement the well-traveled Kremer-Grest coarse-grained model~\cite{kremer_dynamics_1990} to represent both the polymer matrix and the particles that constitute filler particles in the simulations. This broad survey is conducted within the confines of a single `universal' model design so that trends that span the various parameter sweeps are more clearly correlated with their microscopic origins. 

\section{\label{sec:methods}Computational Model and Methods}
\subsection{\label{sec:model}Force field and model}
Pairwise interactions are described by the Lennard-Jones (L-J) potential, 
\begin{equation}
  U_{ij}(r) = 4\epsilon_{ij}\left[\left(\frac{\sigma_{ij}}{r}\right)^{12}-\left(\frac{\sigma_{ij}}{r}\right)^{6}\right],
\label{eqn:LJ}
\end{equation}
where $r$ is the distance between coarse grained particles $i$ and $j$. To generalize our results, we employ the Lennard-Jones unit system and let $k_BT=\epsilon=1$ for all possible interactions between polymer and filler particle bead types. All particle types are assigned a mass of 1 and have a diameter $\sigma=1$. We use the finite extensible nonlinear elastic (FENE) potential to represent bonds between beads in linear polymer chains,  
\begin{eqnarray}
  U_{b}(r) =&& - \frac{1}{2}k_b r_\text{max}^2 \text{ln}\left[ 1- \left( \frac{r}{r_\text{max}} \right)^2 \right]\nonumber\\
&&+4\epsilon\left[\left(\frac{\sigma}{r}\right)^{12}-\left(\frac{\sigma}{r}\right)^{6}\right],
\label{eqn:FENE}
\end{eqnarray}
where the first term is attractive and contains the force constant $k_b=30$ and $r_\text{max}$ is the maximum bond length, $1.5 \sigma$. The second term is cut off at a distance of $r=2^{1/6}\sigma$, the minimum of the 12-6 L-J potential, making it fully repulsive~\cite{kremer_dynamics_1990}. 

This work considers both canonical, repulsive-only interactions, where the pairwise interactions in Eqn. \ref{eqn:LJ} are cut off at a distance of $r=2^{1/6}\sigma$, and the case of attractive particles, where Eqn. \ref{eqn:LJ} is cut off at $r=1.75\sigma$~\cite{lin_origin_2020,lin_effect_2021}. We also simulate the cases of dissimilar interactions between the polymer and FP; where polymer-polymer and FP-FP interactions are repulsive-only and polymer-FP interactions are attractive and the inverse case where polymer-FP interactions are repulsive-only. All simulations are performed using the open-source LAMMPS codebase~\cite{thompson_lammps_2021}.

Model filler particles (FP) are constructed by arranging K-G beads onto the surface of a sphere. To do so, we define the centers of the surface beads by constructing a Fibonacci Sphere of the desired radius $r$ with $4\pi r^2$ surface points. Each point defines the location of a K-G bead and one additional bead is placed at the center of the FP. Implementation of these composite hollow spheres is considerably less expensive than representing each FP with a single bead of radius $r=10$ or $r=5$, since the latter would dramatically increase the minimum cutoff distance for force calculations in the simulation. Each FP, consisting of $4\pi r^2+1$ beads, is treated as a rigid body which moves and rotates as a single entity during MD simulation. The total force and torque on each FP is computed as the sum of the forces and torques on the constituent beads. 

\subsection{\label{sec:equilib}System setup, relaxation, and mechanical testing}

Composite polymer-particle systems are prepared using our in-house code, which generates a low-density initial configuration based on the desired FP size, FP volume fraction $\phi$, and polymer chain length. We impose a standard final polymer phase density of $0.87\text{ beads}/\sigma^3$ and a final box edge length of $97.31 \sigma$ for all simulations. This set of parameters fully constrains the definition of a polymer-FP system. Initial configurations are prepared by randomly placing the appropriate number of FPs into a box with edge length defined so that the polymer phase will have a density of $0.01 \text{ beads} / \sigma^3$. Polymer strands are then inserted using a self-avoiding random walk approach, with additional considerations taken to ensure that the polymer strands do not overlap the FPs.  These low-density configurations are initially relaxed using a soft interaction potential to escape any high-energy states, followed by introduction of the desired L-J potential. Our relaxation approach is similar to that implemented by Riggleman and co-workers~\cite{lin_origin_2020} and consists of heating the simulation to $T=10$, compression to the final polymer density of $0.87 \text{ beads} / \sigma^3$, and cooling to simulation temperature over the course of $\sim 5 \times 10^6$ MD steps with a timestep $\tau = 0.002$.

\begin{figure}[h]
\includegraphics[width=3.3in]{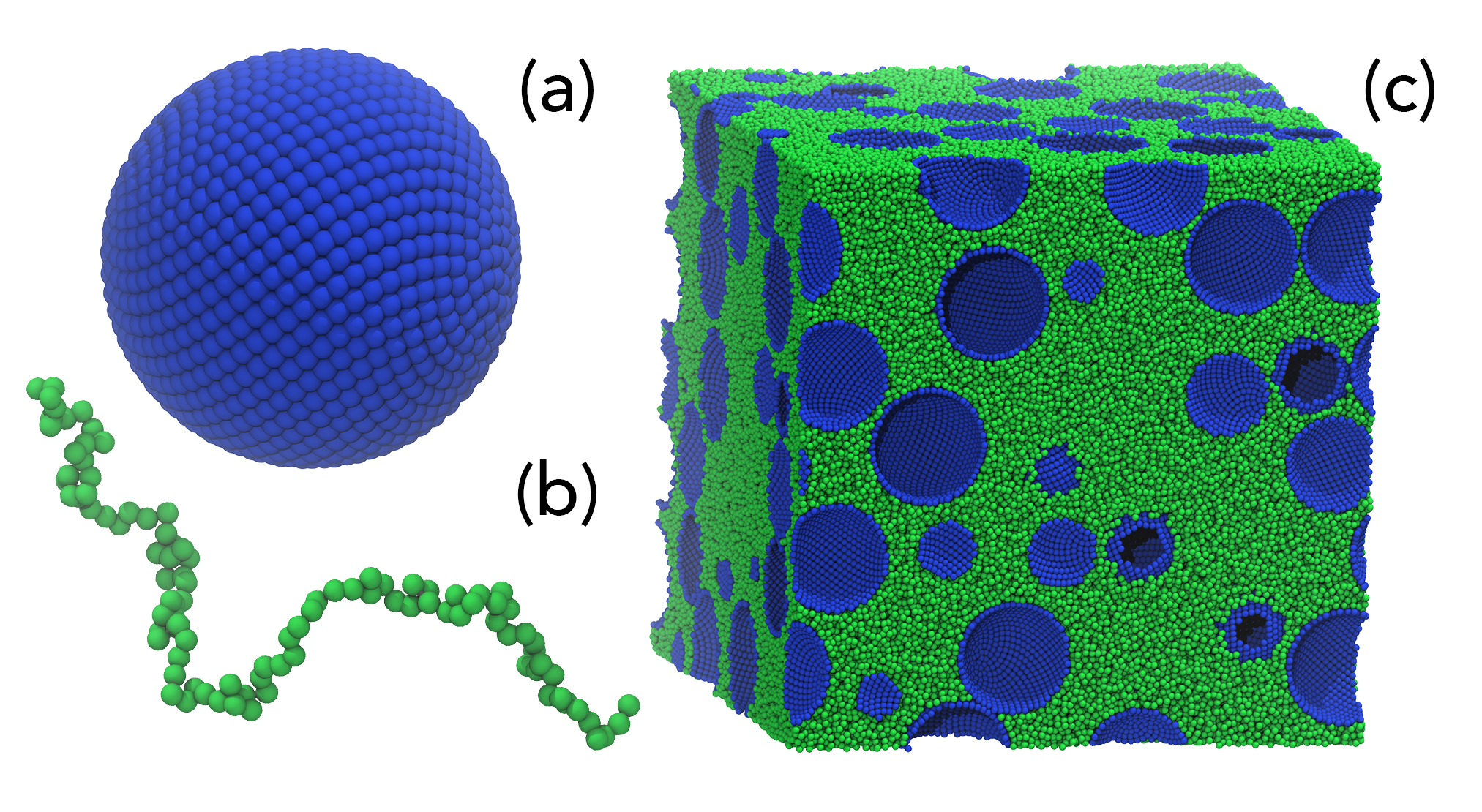}
\caption{\label{fig:box} (a) Hollow Fibonacci spheres of K-G beads, example shown with radius = $10 \sigma$, represent the filler particles and (b) polymer strands of $N=100$ K-G beads are prepared by a self-avoiding random walk. (c) Representative simulation cell with $\phi = 42.5\%$ after high-temperature relaxation and compression to final density.}
\end{figure}

Simulated tensile testing is performed by the uniaxial extension of the cubic simulation box along a selected axis (e.g. $x$) at a constant engineering strain rate of $2 \times 10^{-5}$ while the orthogonal periodic boundaries ($y$ and $z$ axes) are controlled by the Berendsen barostat and therefore allowed to change size while maintaining a constant pressure. To verify isotropic behavior, each relaxed starting configuration was tested by independent elongations along the $x$, $y$, and $z$ axes. All stress-strain data shown in this work represents the average of these three uniaxial extension simulations. We recognize that some observed behavior and phenomena may be functions of strain rate $\dot{\epsilon}$ and present our findings which implement a conservative choice of $\dot{\epsilon}$ and provide insight into both the composite material's mechanical reinforcement and the origins of mechanical failure.

As mentioned earlier, the phase space available to the simulation protocol introduced in this work is vast. We therefore select parameters of interest to vary and explore all resulting permutations. For all simulations, K-G beads have a diameter of $1 \sigma$ and the number of K-G beads per polymer chain is set to $N=100$. In all simulations with filler particles the FP are monodisperse Fibonacci spheres with a radius of $2.5$, $5.0$, or $10.0 \sigma$ and filler volume fractions ($\phi$) ranging from 15\% to 50\%. System temperatures of $T=1.0$ and $T=0.3$ are studied to capture both the canonical K-G model and the low-temperature (glassy) regime recently studied by Riggleman and co-workers~\cite{lin_origin_2020}. Interaction potentials $\epsilon$ may be either fully repulsive, using the Weeks-Chandler-Anderson (WCA) formulation, or allow for inter-particle attraction by extending the cut off of this interaction to $1.75 \sigma$. Polymer-polymer, polymer-filler, and filler-filler interactions are varied, resulting in four scenarios. In this work we refer to the repulsive only interaction as `WCA' and to the attractive potential (i.e. with $r_\text{cutoff}=1.75\sigma$) as `LJ'. Table \ref{tab:param} summarizes simulation settings and the parameter space explored in this work.

The in-house code developed to generate these polymer-composite molecular dynamics starting configurations and LAMMPS input files is freely available at \verb|https://github.com/karnes/K-G-and-Meatballs|
\begin{table}[b]
\caption{\label{tab:param}%
Listing of simulation parameters.
}
\begin{ruledtabular}
\begin{tabular}{lcr}
\textrm{Parameter}&
\textrm{Symbol}&
\textrm{Value(s)}\\
\colrule
Box edge length, $\sigma$ & $l_0$ & 97.31\\
Chain length & $N$ & 100\\
Polymer bulk density, $\sigma^{-3}$ & $\rho$ & 0.87\\
Same-type interaction\footnote{polymer-polymer and FP-FP} & $\epsilon_{ii}$ & WCA, LJ\\
Cross-type interaction\footnote{polymer-FP} & $\epsilon_{ij}$ & WCA, LJ\\
Temperature & $T$ & 0.3, 1.0\\
Filler vol. \% & $\phi$ & 15, 25, 35, 42.5, 50\\
\end{tabular}
\end{ruledtabular}
\end{table}

\section{\label{sec:results}Results and discussion}

\begin{figure}[h]
\includegraphics[width=3.375in]{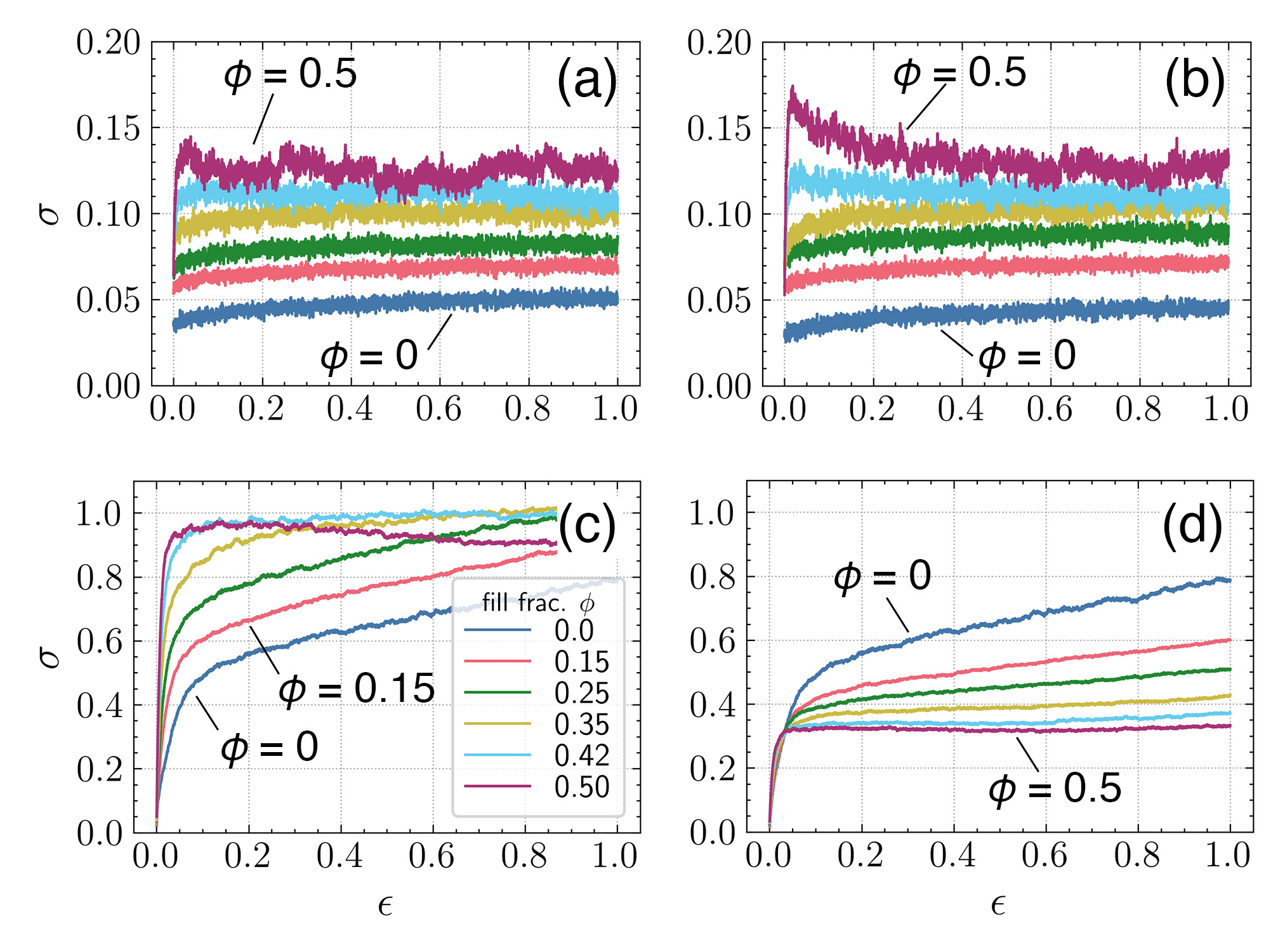}
\caption{Simulated tensile testing results for polymer-filler composites for selected FP fill fraction $\phi$. Panels (a) and (b) are above $T_g$ with WCA (panel (a)) and LJ (panel (b)) interactions implemented between all particle types. Panels (c) and (d) are at $T=0.3$, in the glassy state. In (c) all pairwise interactions are slightly attractive (LJ), while in (d) like interactions ($\epsilon_{ii}$) are LJ and mixed interactions ($\epsilon_{ij}$, i.e. polymer-FP) are described by the repulsive-only WCA potential. All FP have a radius of $10\sigma$.}
\label{fig:strstr-phi}
\end{figure}

Figure \ref{fig:strstr-phi} is an overview of simulated tensile testing as a function of filler particle (FP) volume fraction, $\phi$. In panels (a)-(d) the FP volume percent ranges from $0\% \text{(neat polymer)} < \phi < 50\%$. Stress-strain curves in all panels share the same $\phi$ vs. color scheme as defined in the panel (c) legend. For all cases in Figure \ref{fig:strstr-phi} the radius of the filler particles is $10 \sigma$. We note that, in practice, FP packing above 50 volume percent leads to numerical instabilities during uniaxial extension of the simulation cell, specifically in instances where FENE bond lengths exceed the maximum length, $r>r_\text{max}$  (see Equation \ref{eqn:FENE}). This point of instability is interesting from the physical perspective since it corresponds with the lowest known jammed packing density of $\phi=\pi \sqrt{2}/9 \approx 0.4937$~\cite{torquato_jamming_2007}. This instability can be circumvented by using a harmonic potential to describe bonds within the polymer chain and this approach has been applied to simulate packing fractions up to 60 volume \%~\cite{lin_origin_2020}. For reference, we note that the theoretical limit of close packed spheres is $\phi=\frac{\pi}{3 \sqrt{2}} \approx 0.7405$~\cite{hales_overview_2002}. 

Figure \ref{fig:strstr-phi}(a) shows the stress-strain responses for the canonical K-G model, with the WCA interaction potential and a temperature of $T=1.0$, as a function of $\phi$. The stress increases with increasing $\phi$ for all packing fractions, indicative of the mechanical reinforcement provided by the filler particles. At the highest fill fraction, 0.5, stress strain response is less smooth, suggesting that the simulation box moves through `stick-slip' configurations during the uniaxial extension. Panel (b) shows the response when the attractive LJ potential is implemented. At $T=1.0$ stress-strain response for the WCA and LJ potentials is rather similar, with somewhat increased stress near the early elastic region of the high fill fraction cases. These results confirm that the addition of a potential well that corresponds to $\epsilon=1$ has little impact on the mechanical response of the polymer composite at $T=1$. 

Introduction of this attractive potential results in the polymer phase having a glass transition temperature of $T_g\approx0.4$~\cite{lin_origin_2020}. Panel (c) shows the stress-strain response for the LJ system at $T=0.3$, in the glassy regime. Simulated tensile testing below $T_g$ resulted in stress curves about an order of magnitude larger than the $T=1.0$ cases, with a similar increase in stress with increasing FP packing. At higher packing fractions, yield-like behavior emerges as the polymer domains become smaller and the average FP-FP distances decrease. We note that the glassy regime is only accessible at reduced temperatures when the attractive LJ potential ($\epsilon=1$) is implemented; polymers do not exhibit undergo a glass transition in the canonical K-G model or when interaction strengths are less than 0.4 $k_BT$~\cite{jain_investigation_2004, ford_molecular_2022}. 

Panels (a)-(c) all exhibit increasing stress-strain with increasing $\phi$, regardless of whether the simulation is executed with repulsive-only (WCA) or attractive interparticle potentials (LJ). In panel (d) we mix these interactions to simulate a change in the composite's formulation: like-particle interactions (e.g. polymer-polymer) are LJ but mixed particle interactions, polymer-FP, are represented by the WCA potential to represent FP with chemically passivated surfaces. This reformulation reverses the trend seen in panels (a)-(c), where the non-interacting FP effectively lubricate the material and result in a monatonic decrease in stress-strain response with increasing fill fraction.

\begin{figure}[h]
\includegraphics[width=3.375in]{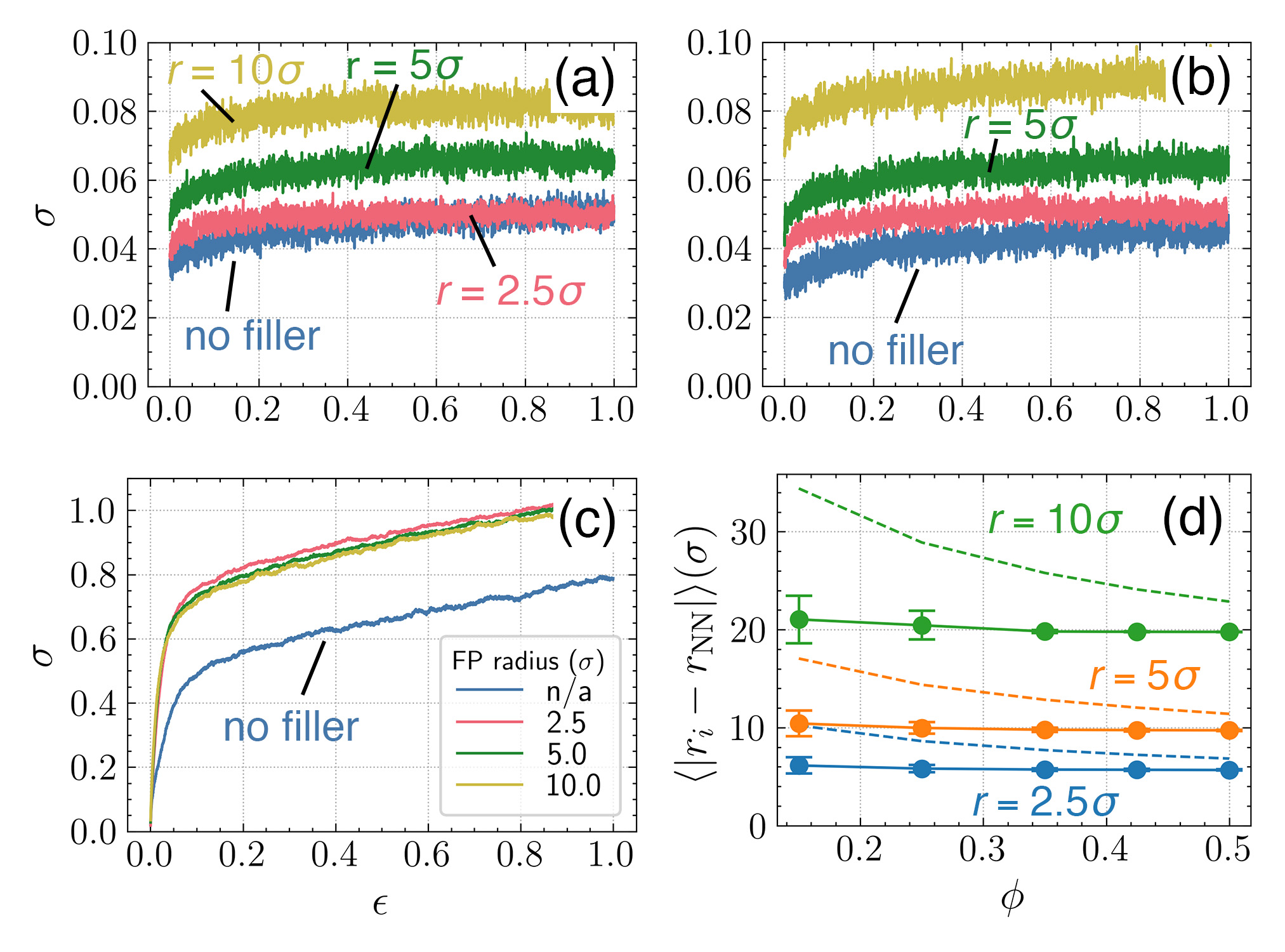}
\caption{Simulated tensile testing as a function of filler particle radius. $\phi=0.25$ for all polymer composites. The top row of panels shows results at $T=1.0$ for the (a) WCA and (b) LJ interaction potentials. Panel (c) shows stress-strain curves for the LJ system at $T=0.3$. In (d), FP nearest neighbor distances from $T=1.0$ WCA simulations are represented by solid curves error bars represent one standard deviation. Values for the corresponding ideal case of equidistant FP are shown as dashed curves.}
\label{fig:strstr-rad}
\end{figure}

Figure \ref{fig:strstr-rad} shows the effect of varying FP radius on tensile response. For these simulations, $\phi=0.25$ is selected as a representative intermediate value of fill fraction. FP radii of $2.5, 5.0,$ and $10.0\sigma$ and the neat polymer system are compared. Similar to the survey of fill fraction $\phi$ in Figure \ref{fig:strstr-phi}, the WCA and LJ potentials in panels \ref{fig:strstr-rad}(a) and (b) are similar to first approximation. At $T=1.0$ stress-strain response increases with increasing FP size, consistent for both force fields. We explain this enhancement by referencing that mechanical reinforcement emerges from the creation of transient polymer networks that bridge the FP. With increased radius of curvature, the bridging networks in larger radius systems are more robust and contribute more enhancement than is provided by the additional number of particles in a smaller FP radius system. When below $T_g$, as in panel (c), the LJ system shows an enhanced mechanical response that is relatively agnostic to FP size. In the glassy state, general rearrangement of the polymer itself during the deformation is much more energetically intensive than the local reordering around the filler particles as a function of FP radius. Although the stress differences are relatively small in the glassy state, systems with larger particle radius provide slightly less enhancement, suggesting that rearranging the glassy polymer around a greater number of smaller FP is more energetically expensive than effects due to local transient reinforcement networks. 

In experimental formulations filler materials similar to those simulated in this work are known to self-assemble into larger, tightly bound agglomerates, often with dendritic morphology~\cite{jouault_WellDispersed_2009}. We did not implement forces to explicitly encourage or discourage the agglomeration of filler particles but note that this behavior was observed in all simulations. Figure \ref{fig:strstr-phi}(d) attempts to summarize the global drive toward agglomeration observed in our simulations. The average FP-FP nearest neighbor distance, $\langle | r_i - r_{\text{NN}} | \rangle$, where $r_i$ is the position of an FP of interest and $r_{\text{NN}}$ is the position of its nearest neighbor FP, is plotted versus $\phi$ for selected FP radii. For reference, we include dashed curves which represent the FP-FP separation distance for the ideal case of maximum separation, where FP are equidistant from each other within the polymer. For all cases, the value of $\langle | r_i - r_{\text{NN}} | \rangle$ is close to twice the FP radius, deviating significantly from the case of equal spacing. The pure repulsive WCA case is shown in Figure 4(d), highlighting the entropic drive toward particle agglomeration. We reserve further investigation of the effects of FP spacing, FP-FP bonding, and FP agglomerate morphology for future work.  

\begin{figure}[h]
\includegraphics[width=3.3in]{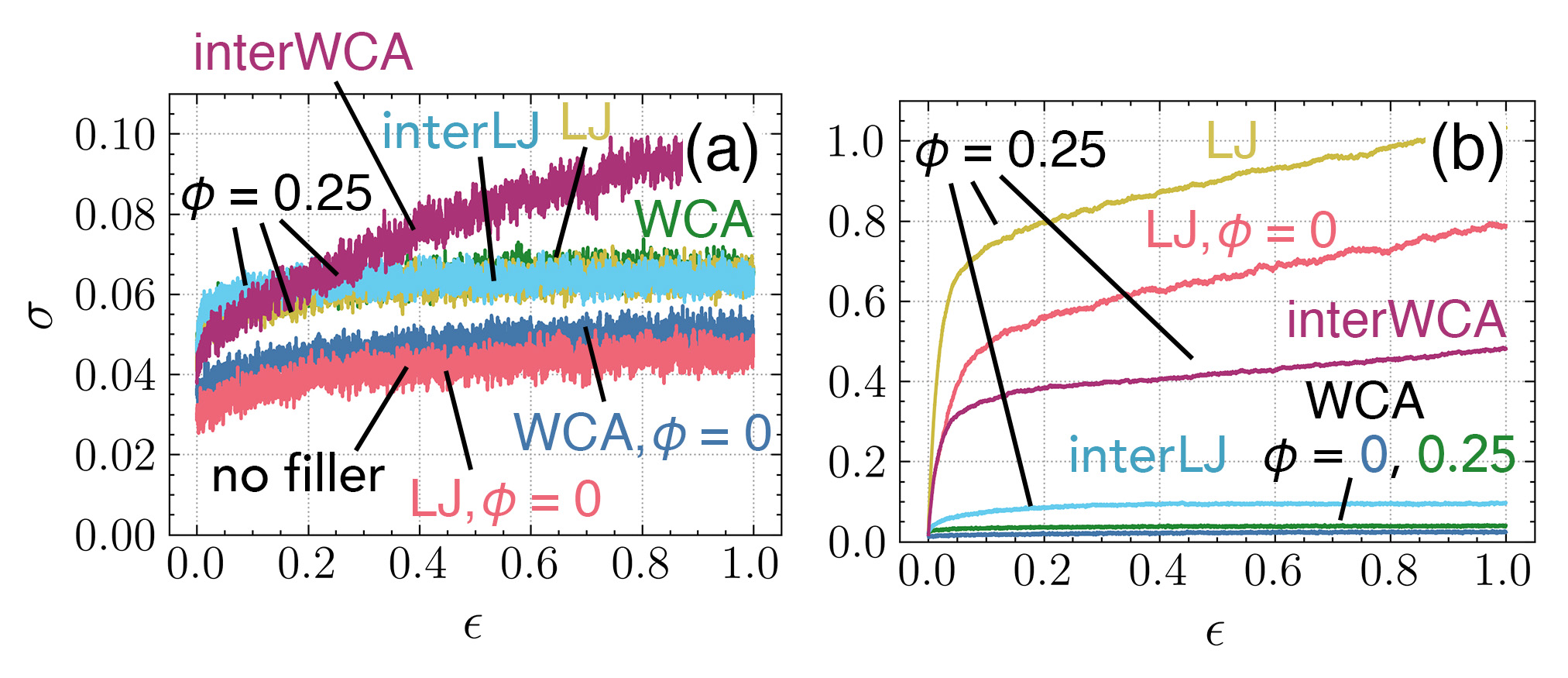}
\caption{Simulated stress-strain response as a function of system interaction potential for (a) $T=1.0$ and (b) $T=0.3$ conditions. For all composite systems FP radius $r=10\sigma$.}
\label{fig:strstr-eps} 
\end{figure}

Figure \ref{fig:strstr-eps} further considers the effects of interaction potential on the systems' stress-strain response. In panel (a) the temperature was held at $T=1.0$ and both the neat polymer and 25 FP volume \% configurations were considered, $\phi=0, 0.25$. As observed earlier, the difference between WCA and LJ interaction potentials is very small for the neat polymer and $\phi=0.25$ systems. Here we also introduce mixed interaction potentials to simulate dissimilar FP and polymer chemistries, which is to be expected in most laboratory formulations. For this approach the first of the two cases (introduced but not named in Figure \ref{fig:strstr-phi}(d)) implements LJ interactions between like particles and WCA interaction between dissimilar particles: Polymer-polymer interactions and FP-FP interactions have an attractive LJ potential with $\epsilon=1$ and polymer-FP interactions use the WCA potential, which contains only the repulsive part of the Lennard-Jones potential. We refer to this case as $interWCA$ to indicate that the interactions between dissimilar particle types are governed by the WCA potential. The second case reverses the interactions: polymer-polymer and FP-FP use the WCA potential and polymer-FP interactions are the attractive LJ. We refer to this case as $interLJ$.

The $interLJ$ system behaves similarly to the pure WCA and LJ cases in Figure \ref{fig:strstr-eps}(a), which indicates that introducing an attraction between polymer and FP does not significantly perturb system behavior at $T=1.0$. The $interWCA$ potential does introduce new behavior, showing increasing stress with deformation. At the microscopic level, deformation of the $interWCA$ system appears to create behavior analogous to the hydrophobic effect, where disruption of the cohesive polymer phase by filler particles (which are similarly attracted to other filler particles) is energetically unfavorable in addition to being entropically unfavorable as in the WCA and LJ cases.  

Tensile responses become more complex at $T=0.3$ in Figure \ref{fig:strstr-eps}(b). Both of the WCA systems, $\phi=0$ and $\phi=0.25$, behave similarly to $T=1.0$ since these configurations are not in the glassy state. The interaction potentials used in the $interLJ$ system are similar to the WCA $\phi=0.25$ system but introduce attraction between FP and polymer, resulting in a noticeably higher stress during deformation. This additional stress is notable in the context of designing composite materials since it represents both preferential polymer-filler interactions and the tethering of the polymer to the FPs. These interactions, although weak, reduce polymer mobility when near the FPs and result in noticeable mechanical enhancement. The increased stress due to the $interLJ$ potential is a small perturbation compared to the neat polymer at $T=0.3$ with LJ interaction. The tensile response of the neat glassy polymer is a convenient baseline for comparing the final two composite systems, which add 25 volume \% filler to the system. Addition of filler particles with the LJ potential mechanically reinforces the system, with behavior seen in Figure \ref{fig:strstr-phi}(c). In the $interWCA$ case, the FP-polymer interface is governed by the repulsive-only WCA potential and the FPs effectively lubricate motion in the glassy system, resulting in a more easily deformed system: not the mechanical enhancement expected by the addition of FPs. This behavior has important implications for the designer of systems where the composite material will be exposed to temperatures below the $T_g$ of the polymer.

\begin{figure}[h]
\includegraphics[width=3.3in]{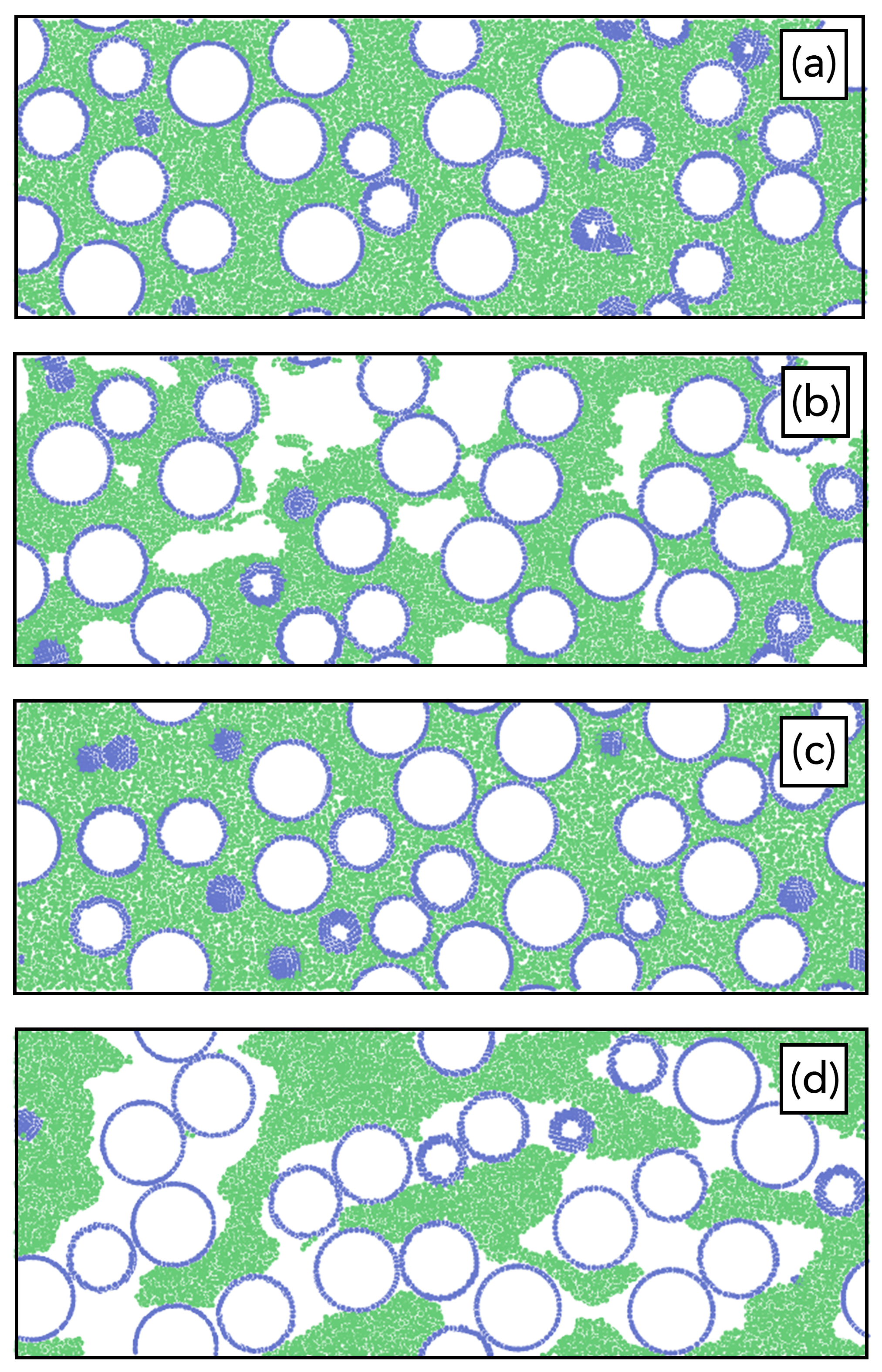}
\caption{Simulation snapshots of 100\% elongated polymer composites. Each snapshot is a 2-D slice parallel to the axis of uniaxial extension. The respective interaction potentials are (a) WCA, (b) LJ, (c) $interLJ$, and (d)$interWCA$.}
\label{fig:delam}
\end{figure}

Beyond mechanical stress-strain response, the potential for material failure must be considered when designing or formulating composite materials. In this and earlier work, inspection of MD configurations reveals the emergence of voids within the deformed materials. Figure \ref{fig:delam} shows simulation snapshots acquired at $T=0.3$ and 100\% strain for the $\phi=0.425$ system and includes cases where significant delamination and voids have formed within the deposit. Each snapshot is a 2-D slice of the simulation cell, parallel to the axis of elongation. Polymer beads are green, FP particles are blue and the hollows within the rigid FP Fibonacci spheres are left uncolored. Panel (a) was obtained from the WCA system and no voidspace within the polymer phase is observed. The LJ potential was implemented in panel (b) and the glassy polymer composite has a noticeable population of voids within the polymer phase. These voids are located both adjacent to the FPs and isolated within the polymer. This interfacial/intra-polymer distribution is reasonable and anticipated since the polymer-polymer and polymer-FP interaction potentials are identical in this simulation. Panel (c) shows the case of $interLJ$, where the polymer-polymer interaction is not attractive but the polymer-FP interaction is. No voids appear in this case; the WCA polymer phase is well above its $T_g$ and the small glassy domains at the polymer/FP interface do not introduce a noticable morphological perturbation during the simulated tensile testing. In panel (d) voids are present and appear to be exclusively located at the polymer/FP interfaces. The $interWCA$ potential implemented in panel (d) consists of attractive polymer-polymer interactions, and therefore a glassy phase at $T=0.3$. The polymer-FP interactions are WCA, which includes only a repulsive component, thus making the polymer/FP interface a logical point of origin for delamination to occur during mechanical deformation. The snapshots in Figure \ref{fig:delam} are representative of system behavior at $T=0.3$, but this mode of analysis is inherently qualitative.

We use PoreBlazer 4.0~\cite{sarkisov_materials_2020} to quantify the sizes of the voids formed during simulated tensile testing. In brief, PoreBlazer computes a Pore Size Distribution (PSD) by selecting random points within the simulation that do not overlap the K-G beads. Each point is the center of a sphere and assigned a maximum diameter possible without overlapping any of the beads within the simulation. We refer the interested reader to the original text for a full description of the PSD algorithm~\cite{sarkisov_materials_2020}.

\begin{figure}[h]
\includegraphics[width=3.3in]{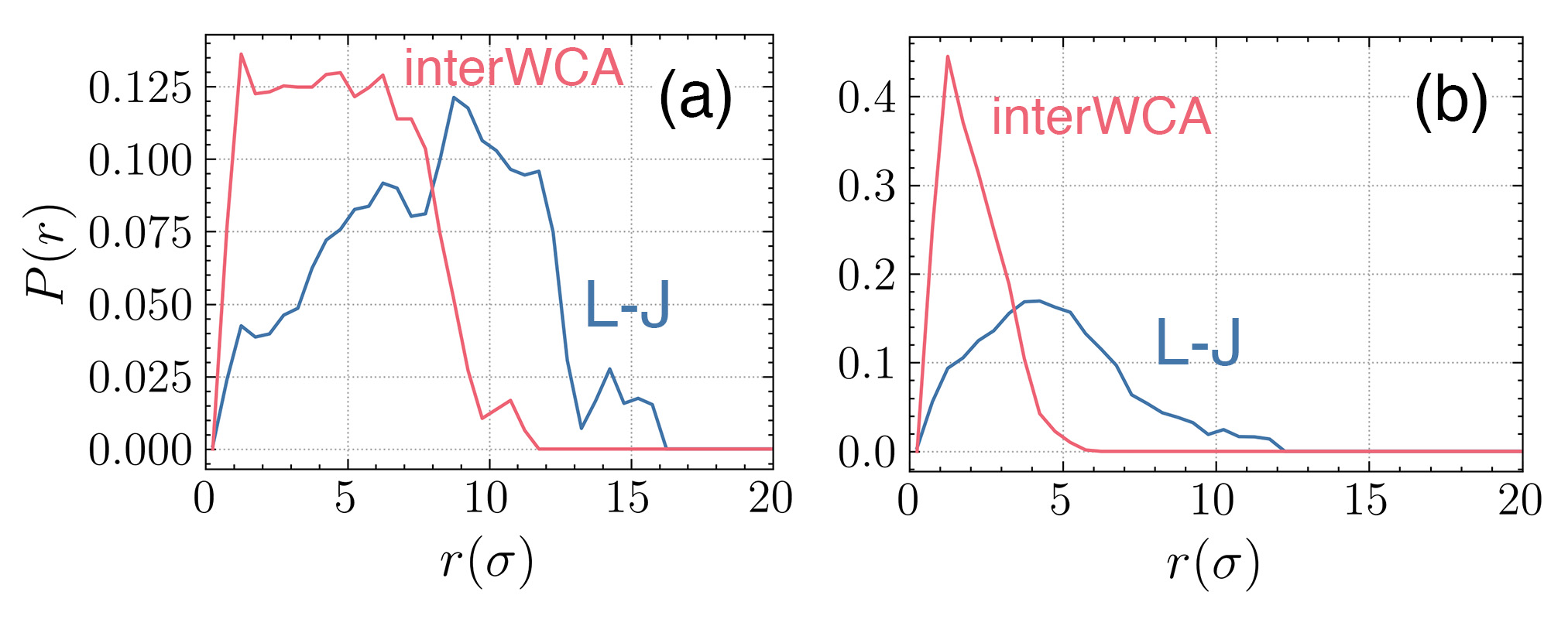}
\caption{Pore distribution functions quantify the voids that emerge during simulated tensile testing of the polymer composites. Panel (a) shows the case of $r=10\sigma$ FP, (b) shows the distributions for $r=5\sigma$.}
\label{fig:pb} 
\end{figure}

Figure \ref{fig:pb} (a) shows PSDs which correspond to the snapshots in Figure \ref{fig:delam} (b) and (d), with $\phi=0.5$ and FP radius $r=10\sigma$ for both cases. For the $interWCA$ potential, where void formation initiates at the polymer/FP interface, the pore sizes are generally smaller than in the LJ simulation, where voids form at any given location. This behavior is consistent for other FP sizes as shown in Figure \ref{fig:pb}(b). These simulations are similar to panel (a), with $\phi=0.5$ but a smaller FP radius of $r=5\sigma$. For both sizes of FP, the average void radius is approximately equal to half of the FP radius for the $interWCA$ potential and slightly less than the FP radius for the LJ potential. The voids may appear larger in panel (d) of Figure \ref{fig:delam} than in panel (b) but the actual sizes are smaller since, in most cases, filler particles are located within the void spaces. PSDs are useful for quantitatively describing the size of the voids within the polymer composites but they do not provide complementary information regarding the spatial location of these voids.

A cursory inspection of Figure \ref{fig:delam} (b) and (d) suggests that voids form at the polymer/FP interface in the $interWCA$ case during uniaxial extension. To quantify this behavior, we developed an approach based on the standard pair distribution function $g_{ij}(r)$ calculation where, in this work, $i$ is the center point of the filler particles and $j$ represents and arbitrary point located within the voidspace. To implement, we randomly inserted particles of a new type, diameter $1\sigma$ but dissimilar to both the FP and polymer beads into the elongated configuration until all void space was filled with these `void' particles. This set of void particles serves to discretize the voids and enable quantitative description of their position relative to the FP by familiar means. From here, the calculation of a `voidspace' distribution function, $g_v(r)$, is straightforward,

\begin{equation}
  g_v(r)=(N_\text{VP}N_\text{FP})^{-1} \sum_{i=1}^{N_\text{VP}}\sum_{j=1}^{N_\text{FP}}\langle \delta ( | \textbf{r}_i-\textbf{r}_j | - r ) \rangle
  \label{eqn:gvr}
\end{equation}

where VP and FP represent the void particles and centerpoints of the filler particles. A complete description of voidspace distribution function algorithm and its implementation is included in the Supplementary Material.

Figure \ref{fig:gr} shows the voidspace distribution functions for $\phi=0.5$ and FP radius $r=10\sigma$. The region $0<r<10\sigma$ where $g_v(r)\approx0$ corresponds with the volume excluded by the filler particles. The sharp peak in the $interWCA$ curve at $r=11\sigma$ and associated shoulder correspond with a large population of voidspace adjacent to the FP particles' surfaces. This is contrasted by the LJ curve, which shows a small shoulder peak at the FP surface followed by a broad peak centered at about $1.5 \times$ the FP radius. 

\begin{figure}[h]
\includegraphics[width=3.3in]{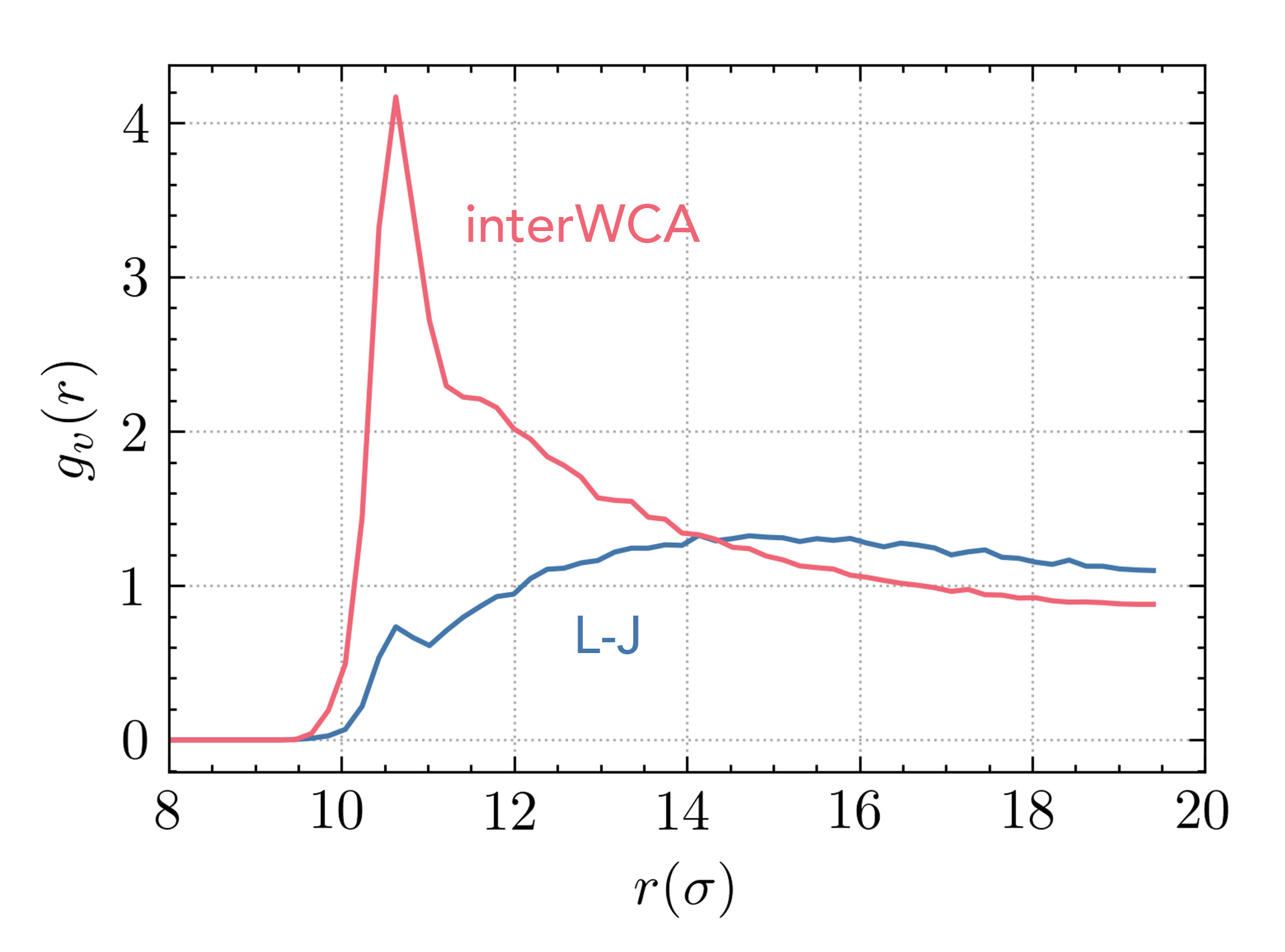}
\caption{Voidspace distribution functions $g_v(r)$ quantify the location of voids relative to the filler particle centers for the LJ and $interWCA$ potential simulations.}
\label{fig:gr}
\end{figure}

\begin{figure}[h]
\includegraphics[width=3.3in]{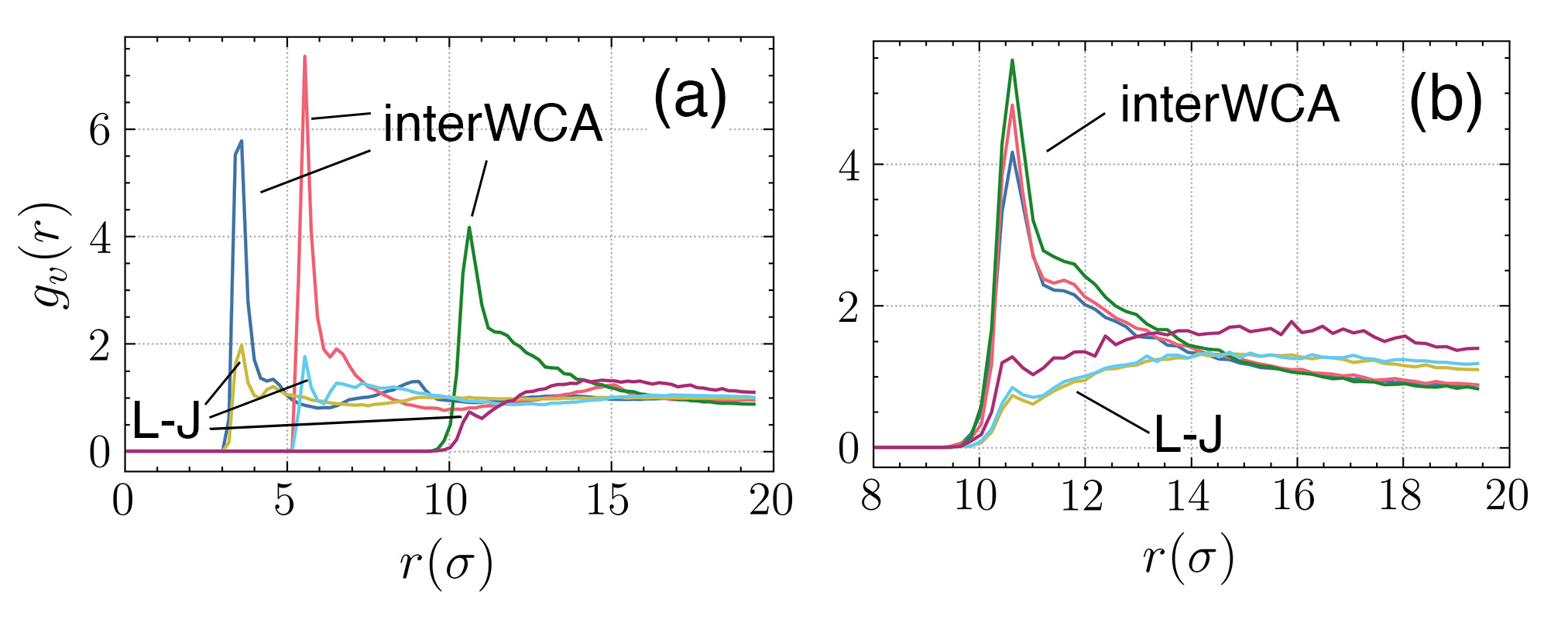}
\caption{Voidspace distribution functions for (a) different FP radii and (b) fill fractions $\phi$ reveal the consistent spatial arrangement of cavitation and delamination during simulated uniaxial extension of the polymer composites.}
\label{fig:gr-rad} 
\end{figure}

Figure \ref{fig:gr-rad} (a) extends this analysis for different FP radii and confirms that the behavior is consistent. For all FP sizes surveyed in this work the the weaker polymer-FP interaction in the \textit{interWCA} potential results in more pronounced delamination at the FP interface than in the LJ case, where the attractive potential is constant for all pairwise species interactions. Panel (b) considers the void distribution function $g_v(r)$ for a range of fill fractions $\phi$ = 0.35, 0.425, and 0.5 and FP radius $10 \sigma$. The voids formed during simulated uniaxial extension are located at the FP-polymer interface for all \textit{interWCA} cases as indicated by the sharp peaks at $r=10\sigma$. For the \textit{LJ} simulations the distribution function is more broad, with a relatively small peak at $10\sigma$, indicating that the origin of each delamination is largely agnostic to the presence of FP-polymer interfaces. For all $\phi$ the respective curves for the \textit{interWCA} and \textit{LJ} cases look very similar to first approximation. The peak heights for \textit{interWCA} increase with decreasing $\phi$, which we attribute to more bridging of the voids between FP and increased numbers of local FP-FP neighbors reducing the magnitude of $g_v(r)$. A similar phenomena is seen in the \textit{LJ} case, where the curves increase with decreasing $\phi$ since the lower packing fractions permit the formation of larger voids.


\section{\label{sec:conclusions}Conclusions}

We implemented the well-traveled Kremer-Grest bead-spring representation of a polymer system to simulate model polymer-filler composites. Varying the interaction potentials enables a unique and rapid exploration of design phase space that allows researchers to rapidly survey the impact of high-level design choices, such as the interaction strength between filler material and bulk polymer. Further, these simulations circumvent the costly formulate-synthesize-cure-test benchtop development cycle and, due to the generality of the model, decouple the design of these materials from the selection of backbone and functional group chemistry. This high-level approach to polymer interrogation separates the design from limitations of standard feedstock and synthetic technique and, when a promising candidate is found, can encourage polymer scientists to seek out methods to realize these designs.

Varying FP packing fraction and diameter affect the mechanical enhancement provided by the filler and our simulations suggest that non-trivial jammed states emerge and influence stress-strain behavior at higher $\phi$. Introducing FP with passivated surfaces, i.e. treated to minimize the interaction potential between the FP and surrounding polymer, was shown to invert the trend seen in systems with a homogeneous interparticle potential. When $\epsilon_{ii}=\epsilon_{ij}$ a monotonic increase in mechanical reinforcement with increasing $\phi$ is observed. Introducing noninteracting, `passivated' FP into a cohesive reverses this relationship.

In the glassy state, voids or delaminations may emerge during deformation of the polymer composite. Analysis of the pore sizes revealed distinct differences as a function of the polymer-FP interaction potential: When the particles and polymer interact with the same attractive potential, the voids formed are significantly larger than when the polymer is not attracted to the FP. To identify the location of these voids relative to the morphology of the composite, we introduced the void distribution function $g_v(r)$, which combines a controlled insertion of dummy particles with standard pair distribution function. This analysis allows for quantification of voidspace relative to the filler particles and shows that, in the case of passivated FP delamination is much more likely to occur at the polymer-FP interface than within the glassy polymer matrix. 

\begin{acknowledgments}
This work was performed under the auspices of the U.S. Department of Energy by Lawrence Livermore National Laboratory under Contract DE-AC52-07NA27344, release number LLNL-JRNL-2003849. The authors thank S. Schmidt, J. M. Lenhardt, J. A. Armas, J. R. Gissinger, and L. S. Thornbury for many helpful conversations. 
\end{acknowledgments}

\bibliography{ref}

\begin{thebibliography}{38}%
\makeatletter
\providecommand \@ifxundefined [1]{%
 \@ifx{#1\undefined}
}%
\providecommand \@ifnum [1]{%
 \ifnum #1\expandafter \@firstoftwo
 \else \expandafter \@secondoftwo
 \fi
}%
\providecommand \@ifx [1]{%
 \ifx #1\expandafter \@firstoftwo
 \else \expandafter \@secondoftwo
 \fi
}%
\providecommand \natexlab [1]{#1}%
\providecommand \enquote  [1]{``#1''}%
\providecommand \bibnamefont  [1]{#1}%
\providecommand \bibfnamefont [1]{#1}%
\providecommand \citenamefont [1]{#1}%
\providecommand \href@noop [0]{\@secondoftwo}%
\providecommand \href [0]{\begingroup \@sanitize@url \@href}%
\providecommand \@href[1]{\@@startlink{#1}\@@href}%
\providecommand \@@href[1]{\endgroup#1\@@endlink}%
\providecommand \@sanitize@url [0]{\catcode `\\12\catcode `\$12\catcode `\&12\catcode `\#12\catcode `\^12\catcode `\_12\catcode `\%12\relax}%
\providecommand \@@startlink[1]{}%
\providecommand \@@endlink[0]{}%
\providecommand \url  [0]{\begingroup\@sanitize@url \@url }%
\providecommand \@url [1]{\endgroup\@href {#1}{\urlprefix }}%
\providecommand \urlprefix  [0]{URL }%
\providecommand \Eprint [0]{\href }%
\providecommand \doibase [0]{https://doi.org/}%
\providecommand \selectlanguage [0]{\@gobble}%
\providecommand \bibinfo  [0]{\@secondoftwo}%
\providecommand \bibfield  [0]{\@secondoftwo}%
\providecommand \translation [1]{[#1]}%
\providecommand \BibitemOpen [0]{}%
\providecommand \bibitemStop [0]{}%
\providecommand \bibitemNoStop [0]{.\EOS\space}%
\providecommand \EOS [0]{\spacefactor3000\relax}%
\providecommand \BibitemShut  [1]{\csname bibitem#1\endcsname}%
\let\auto@bib@innerbib\@empty
\bibitem [{\citenamefont {Staudinger}(1920)}]{staudinger_uber_1920}%
  \BibitemOpen
  \bibfield  {author} {\bibinfo {author} {\bibfnamefont {H.}~\bibnamefont {Staudinger}},\ }\bibfield  {title} {\bibinfo {title} {{\"U}ber {{Polymerisation}}},\ }\href {https://doi.org/10.1002/cber.19200530627} {\bibfield  {journal} {\bibinfo  {journal} {Berichte der deutschen chemischen Gesellschaft (A and B Series)}\ }\textbf {\bibinfo {volume} {53}},\ \bibinfo {pages} {1073} (\bibinfo {year} {1920})}\BibitemShut {NoStop}%
\bibitem [{\citenamefont {M{\"u}lhaupt}(2004)}]{mulhaupt_hermann_2004}%
  \BibitemOpen
  \bibfield  {author} {\bibinfo {author} {\bibfnamefont {R.}~\bibnamefont {M{\"u}lhaupt}},\ }\bibfield  {title} {\bibinfo {title} {Hermann {{Staudinger}} and the {{Origin}} of {{Macromolecular Chemistry}}},\ }\href {https://doi.org/10.1002/anie.200330070} {\bibfield  {journal} {\bibinfo  {journal} {Angewandte Chemie International Edition}\ }\textbf {\bibinfo {volume} {43}},\ \bibinfo {pages} {1054} (\bibinfo {year} {2004})}\BibitemShut {NoStop}%
\bibitem [{\citenamefont {Rubinstein}(2003)}]{rubinstein_polymer_2003}%
  \BibitemOpen
  \bibfield  {author} {\bibinfo {author} {\bibfnamefont {M.}~\bibnamefont {Rubinstein}},\ }\href@noop {} {\emph {\bibinfo {title} {Polymer {{Physics}}}}}\ (\bibinfo  {publisher} {Oxford University Press},\ \bibinfo {year} {2003})\BibitemShut {NoStop}%
\bibitem [{\citenamefont {Tillotson}\ and\ \citenamefont {Hrubesh}(1992)}]{tillotson_transparent_1992}%
  \BibitemOpen
  \bibfield  {author} {\bibinfo {author} {\bibfnamefont {T.}~\bibnamefont {Tillotson}}\ and\ \bibinfo {author} {\bibfnamefont {L.}~\bibnamefont {Hrubesh}},\ }\bibfield  {title} {\bibinfo {title} {Transparent ultralow-density silica aerogels prepared by a two-step sol-gel process},\ }\href {https://doi.org/10.1016/S0022-3093(05)80427-2} {\bibfield  {journal} {\bibinfo  {journal} {Journal of Non-Crystalline Solids}\ }\textbf {\bibinfo {volume} {145}},\ \bibinfo {pages} {44} (\bibinfo {year} {1992})}\BibitemShut {NoStop}%
\bibitem [{\citenamefont {Pekala}\ and\ \citenamefont {Alviso}(1992)}]{pekala_carbon_1992}%
  \BibitemOpen
  \bibfield  {author} {\bibinfo {author} {\bibfnamefont {R.~W.}\ \bibnamefont {Pekala}}\ and\ \bibinfo {author} {\bibfnamefont {C.~T.}\ \bibnamefont {Alviso}},\ }\bibfield  {title} {\bibinfo {title} {Carbon {{Aerogels}} and {{Xerogels}}},\ }\href {https://doi.org/10.1557/PROC-270-3} {\bibfield  {journal} {\bibinfo  {journal} {MRS Online Proceedings Library}\ }\textbf {\bibinfo {volume} {270}},\ \bibinfo {pages} {3} (\bibinfo {year} {1992})}\BibitemShut {NoStop}%
\bibitem [{\citenamefont {Pekala}\ and\ \citenamefont {Schaefer}(1993)}]{pekala_structure_1993}%
  \BibitemOpen
  \bibfield  {author} {\bibinfo {author} {\bibfnamefont {R.~W.}\ \bibnamefont {Pekala}}\ and\ \bibinfo {author} {\bibfnamefont {D.~W.}\ \bibnamefont {Schaefer}},\ }\bibfield  {title} {\bibinfo {title} {Structure of organic aerogels. 1. {{Morphology}} and scaling},\ }\href {https://doi.org/10.1021/ma00072a029} {\bibfield  {journal} {\bibinfo  {journal} {Macromolecules}\ }\textbf {\bibinfo {volume} {26}},\ \bibinfo {pages} {5487} (\bibinfo {year} {1993})}\BibitemShut {NoStop}%
\bibitem [{\citenamefont {Witten}\ \emph {et~al.}(1993)\citenamefont {Witten}, \citenamefont {Rubinstein},\ and\ \citenamefont {Colby}}]{witten_reinforcement_1993}%
  \BibitemOpen
  \bibfield  {author} {\bibinfo {author} {\bibfnamefont {T.~A.}\ \bibnamefont {Witten}}, \bibinfo {author} {\bibfnamefont {M.}~\bibnamefont {Rubinstein}},\ and\ \bibinfo {author} {\bibfnamefont {R.~H.}\ \bibnamefont {Colby}},\ }\bibfield  {title} {\bibinfo {title} {Reinforcement of rubber by fractal aggregates},\ }\href {https://doi.org/10.1051/jp2:1993138} {\bibfield  {journal} {\bibinfo  {journal} {Journal de Physique II}\ }\textbf {\bibinfo {volume} {3}},\ \bibinfo {pages} {367} (\bibinfo {year} {1993})}\BibitemShut {NoStop}%
\bibitem [{\citenamefont {Yadav}\ \emph {et~al.}(2023)\citenamefont {Yadav}, \citenamefont {Singh}, \citenamefont {Shekhawat}, \citenamefont {Lee},\ and\ \citenamefont {Park}}]{yadav_role_2023}%
  \BibitemOpen
  \bibfield  {author} {\bibinfo {author} {\bibfnamefont {R.}~\bibnamefont {Yadav}}, \bibinfo {author} {\bibfnamefont {M.}~\bibnamefont {Singh}}, \bibinfo {author} {\bibfnamefont {D.}~\bibnamefont {Shekhawat}}, \bibinfo {author} {\bibfnamefont {S.-Y.}\ \bibnamefont {Lee}},\ and\ \bibinfo {author} {\bibfnamefont {S.-J.}\ \bibnamefont {Park}},\ }\bibfield  {title} {\bibinfo {title} {The role of fillers to enhance the mechanical, thermal, and wear characteristics of polymer composite materials: {{A}} review},\ }\href {https://doi.org/10.1016/j.compositesa.2023.107775} {\bibfield  {journal} {\bibinfo  {journal} {Composites Part A: Applied Science and Manufacturing}\ }\textbf {\bibinfo {volume} {175}},\ \bibinfo {pages} {107775} (\bibinfo {year} {2023})}\BibitemShut {NoStop}%
\bibitem [{\citenamefont {Jouault}\ \emph {et~al.}(2009)\citenamefont {Jouault}, \citenamefont {Vallat}, \citenamefont {Dalmas}, \citenamefont {Said}, \citenamefont {Jestin},\ and\ \citenamefont {Bou{\'e}}}]{jouault_WellDispersed_2009}%
  \BibitemOpen
  \bibfield  {author} {\bibinfo {author} {\bibfnamefont {N.}~\bibnamefont {Jouault}}, \bibinfo {author} {\bibfnamefont {P.}~\bibnamefont {Vallat}}, \bibinfo {author} {\bibfnamefont {F.}~\bibnamefont {Dalmas}}, \bibinfo {author} {\bibfnamefont {S.}~\bibnamefont {Said}}, \bibinfo {author} {\bibfnamefont {J.}~\bibnamefont {Jestin}},\ and\ \bibinfo {author} {\bibfnamefont {F.}~\bibnamefont {Bou{\'e}}},\ }\bibfield  {title} {\bibinfo {title} {Well-{{Dispersed Fractal Aggregates}} as {{Filler}} in {{Polymer}}-{{Silica Nanocomposites}}: {{Long-Range Effects}} in {{Rheology}}},\ }\href {https://doi.org/10.1021/ma801908u} {\bibfield  {journal} {\bibinfo  {journal} {Macromolecules}\ }\textbf {\bibinfo {volume} {42}},\ \bibinfo {pages} {2031} (\bibinfo {year} {2009})}\BibitemShut {NoStop}%
\bibitem [{\citenamefont {Maiti}\ \emph {et~al.}(2014)\citenamefont {Maiti}, \citenamefont {Small}, \citenamefont {Gee}, \citenamefont {Weisgraber}, \citenamefont {Chinn}, \citenamefont {Wilson},\ and\ \citenamefont {Maxwell}}]{maiti_mullins_2014}%
  \BibitemOpen
  \bibfield  {author} {\bibinfo {author} {\bibfnamefont {A.}~\bibnamefont {Maiti}}, \bibinfo {author} {\bibfnamefont {W.}~\bibnamefont {Small}}, \bibinfo {author} {\bibfnamefont {R.~H.}\ \bibnamefont {Gee}}, \bibinfo {author} {\bibfnamefont {T.~H.}\ \bibnamefont {Weisgraber}}, \bibinfo {author} {\bibfnamefont {S.~C.}\ \bibnamefont {Chinn}}, \bibinfo {author} {\bibfnamefont {T.~S.}\ \bibnamefont {Wilson}},\ and\ \bibinfo {author} {\bibfnamefont {R.~S.}\ \bibnamefont {Maxwell}},\ }\bibfield  {title} {\bibinfo {title} {Mullins effect in a filled elastomer under uniaxial tension},\ }\href {https://doi.org/10.1103/PhysRevE.89.012602} {\bibfield  {journal} {\bibinfo  {journal} {Physical Review E}\ }\textbf {\bibinfo {volume} {89}},\ \bibinfo {pages} {012602} (\bibinfo {year} {2014})}\BibitemShut {NoStop}%
\bibitem [{\citenamefont {Rueda}\ \emph {et~al.}(2017)\citenamefont {Rueda}, \citenamefont {Auscher}, \citenamefont {Fulchiron}, \citenamefont {P{\'e}ri{\'e}}, \citenamefont {Martin}, \citenamefont {Sonntag},\ and\ \citenamefont {Cassagnau}}]{rueda_rheology_2017}%
  \BibitemOpen
  \bibfield  {author} {\bibinfo {author} {\bibfnamefont {M.~M.}\ \bibnamefont {Rueda}}, \bibinfo {author} {\bibfnamefont {M.-C.}\ \bibnamefont {Auscher}}, \bibinfo {author} {\bibfnamefont {R.}~\bibnamefont {Fulchiron}}, \bibinfo {author} {\bibfnamefont {T.}~\bibnamefont {P{\'e}ri{\'e}}}, \bibinfo {author} {\bibfnamefont {G.}~\bibnamefont {Martin}}, \bibinfo {author} {\bibfnamefont {P.}~\bibnamefont {Sonntag}},\ and\ \bibinfo {author} {\bibfnamefont {P.}~\bibnamefont {Cassagnau}},\ }\bibfield  {title} {\bibinfo {title} {Rheology and applications of highly filled polymers: {{A}} review of current understanding},\ }\href {https://doi.org/10.1016/j.progpolymsci.2016.12.007} {\bibfield  {journal} {\bibinfo  {journal} {Progress in Polymer Science}\ }\bibinfo {series} {Topical {{Volume}} on {{Polymer Physics}}},\ \textbf {\bibinfo {volume} {66}},\ \bibinfo {pages} {22} (\bibinfo {year} {2017})}\BibitemShut {NoStop}%
\bibitem [{\citenamefont {Domurath}\ \emph {et~al.}(2017)\citenamefont {Domurath}, \citenamefont {Saphiannikova},\ and\ \citenamefont {Heinrich}}]{domurath_concept_2017}%
  \BibitemOpen
  \bibfield  {author} {\bibinfo {author} {\bibfnamefont {J.}~\bibnamefont {Domurath}}, \bibinfo {author} {\bibfnamefont {M.}~\bibnamefont {Saphiannikova}},\ and\ \bibinfo {author} {\bibfnamefont {G.}~\bibnamefont {Heinrich}},\ }\bibfield  {title} {\bibinfo {title} {The concept of hydrodynamic amplification infilled elastomers},\ }\href@noop {} {\bibfield  {journal} {\bibinfo  {journal} {Kautsch. Gummi Kunstst}\ }\textbf {\bibinfo {volume} {70}},\ \bibinfo {pages} {40} (\bibinfo {year} {2017})}\BibitemShut {NoStop}%
\bibitem [{\citenamefont {Fu}\ \emph {et~al.}(2008)\citenamefont {Fu}, \citenamefont {Feng}, \citenamefont {Lauke},\ and\ \citenamefont {Mai}}]{fu_effects_2008}%
  \BibitemOpen
  \bibfield  {author} {\bibinfo {author} {\bibfnamefont {S.-Y.}\ \bibnamefont {Fu}}, \bibinfo {author} {\bibfnamefont {X.-Q.}\ \bibnamefont {Feng}}, \bibinfo {author} {\bibfnamefont {B.}~\bibnamefont {Lauke}},\ and\ \bibinfo {author} {\bibfnamefont {Y.-W.}\ \bibnamefont {Mai}},\ }\bibfield  {title} {\bibinfo {title} {Effects of particle size, particle/matrix interface adhesion and particle loading on mechanical properties of particulate--polymer composites},\ }\href {https://doi.org/10.1016/j.compositesb.2008.01.002} {\bibfield  {journal} {\bibinfo  {journal} {Composites Part B: Engineering}\ }\textbf {\bibinfo {volume} {39}},\ \bibinfo {pages} {933} (\bibinfo {year} {2008})}\BibitemShut {NoStop}%
\bibitem [{\citenamefont {Vacatello}(2002)}]{vacatello_molecular_2002}%
  \BibitemOpen
  \bibfield  {author} {\bibinfo {author} {\bibfnamefont {M.}~\bibnamefont {Vacatello}},\ }\bibfield  {title} {\bibinfo {title} {Molecular {{Arrangements}} in {{Polymer-Based Nanocomposites}}},\ }\href {https://doi.org/10.1002/1521-3919(20020901)11:7<757::AID-MATS757>3.0.CO;2-I} {\bibfield  {journal} {\bibinfo  {journal} {Macromolecular Theory and Simulations}\ }\textbf {\bibinfo {volume} {11}},\ \bibinfo {pages} {757} (\bibinfo {year} {2002})}\BibitemShut {NoStop}%
\bibitem [{\citenamefont {Lin}\ \emph {et~al.}(2020)\citenamefont {Lin}, \citenamefont {Frischknecht},\ and\ \citenamefont {Riggleman}}]{lin_origin_2020}%
  \BibitemOpen
  \bibfield  {author} {\bibinfo {author} {\bibfnamefont {E.~Y.}\ \bibnamefont {Lin}}, \bibinfo {author} {\bibfnamefont {A.~L.}\ \bibnamefont {Frischknecht}},\ and\ \bibinfo {author} {\bibfnamefont {R.~A.}\ \bibnamefont {Riggleman}},\ }\bibfield  {title} {\bibinfo {title} {Origin of {{Mechanical Enhancement}} in {{Polymer Nanoparticle}} ({{NP}}) {{Composites}} with {{Ultrahigh NP Loading}}},\ }\href {https://doi.org/10.1021/acs.macromol.9b02733} {\bibfield  {journal} {\bibinfo  {journal} {Macromolecules}\ }\textbf {\bibinfo {volume} {53}},\ \bibinfo {pages} {2976} (\bibinfo {year} {2020})}\BibitemShut {NoStop}%
\bibitem [{\citenamefont {Sun}\ \emph {et~al.}(2021)\citenamefont {Sun}, \citenamefont {Melton}, \citenamefont {Safaie}, \citenamefont {Ferrier}, \citenamefont {Cheng}, \citenamefont {Liu}, \citenamefont {Zuo},\ and\ \citenamefont {Wang}}]{sun_molecular_2021}%
  \BibitemOpen
  \bibfield  {author} {\bibinfo {author} {\bibfnamefont {R.}~\bibnamefont {Sun}}, \bibinfo {author} {\bibfnamefont {M.}~\bibnamefont {Melton}}, \bibinfo {author} {\bibfnamefont {N.}~\bibnamefont {Safaie}}, \bibinfo {author} {\bibfnamefont {R.~C.}\ \bibnamefont {Ferrier}}, \bibinfo {author} {\bibfnamefont {S.}~\bibnamefont {Cheng}}, \bibinfo {author} {\bibfnamefont {Y.}~\bibnamefont {Liu}}, \bibinfo {author} {\bibfnamefont {X.}~\bibnamefont {Zuo}},\ and\ \bibinfo {author} {\bibfnamefont {Y.}~\bibnamefont {Wang}},\ }\bibfield  {title} {\bibinfo {title} {Molecular {{View}} on {{Mechanical Reinforcement}} in {{Polymer Nanocomposites}}},\ }\href {https://doi.org/10.1103/PhysRevLett.126.117801} {\bibfield  {journal} {\bibinfo  {journal} {Physical Review Letters}\ }\textbf {\bibinfo {volume} {126}},\ \bibinfo {pages} {117801} (\bibinfo {year} {2021})}\BibitemShut {NoStop}%
\bibitem [{\citenamefont {Shi}\ \emph {et~al.}(2023)\citenamefont {Shi}, \citenamefont {Yu}, \citenamefont {Zhang}, \citenamefont {Yang}, \citenamefont {Lu},\ and\ \citenamefont {Qian}}]{shi_molecular_2023}%
  \BibitemOpen
  \bibfield  {author} {\bibinfo {author} {\bibfnamefont {R.}~\bibnamefont {Shi}}, \bibinfo {author} {\bibfnamefont {L.}~\bibnamefont {Yu}}, \bibinfo {author} {\bibfnamefont {N.}~\bibnamefont {Zhang}}, \bibinfo {author} {\bibfnamefont {Y.}~\bibnamefont {Yang}}, \bibinfo {author} {\bibfnamefont {Z.-Y.}\ \bibnamefont {Lu}},\ and\ \bibinfo {author} {\bibfnamefont {H.-J.}\ \bibnamefont {Qian}},\ }\bibfield  {title} {\bibinfo {title} {Molecular {{Origin}} of the {{Reinforcement Effect}} and {{Its Strain-Rate Dependence}} in {{Polymer Nanocomposite Glass}}},\ }\href {https://doi.org/10.1021/acsmacrolett.3c00235} {\bibfield  {journal} {\bibinfo  {journal} {ACS Macro Letters}\ }\textbf {\bibinfo {volume} {12}},\ \bibinfo {pages} {1052} (\bibinfo {year} {2023})}\BibitemShut {NoStop}%
\bibitem [{\citenamefont {Shen}\ \emph {et~al.}(2020)\citenamefont {Shen}, \citenamefont {Lin}, \citenamefont {Liu},\ and\ \citenamefont {Li}}]{shen_revisiting_2020}%
  \BibitemOpen
  \bibfield  {author} {\bibinfo {author} {\bibfnamefont {J.}~\bibnamefont {Shen}}, \bibinfo {author} {\bibfnamefont {X.}~\bibnamefont {Lin}}, \bibinfo {author} {\bibfnamefont {J.}~\bibnamefont {Liu}},\ and\ \bibinfo {author} {\bibfnamefont {X.}~\bibnamefont {Li}},\ }\bibfield  {title} {\bibinfo {title} {Revisiting stress--strain behavior and mechanical reinforcement of polymer nanocomposites from molecular dynamics simulations},\ }\href {https://doi.org/10.1039/D0CP02225J} {\bibfield  {journal} {\bibinfo  {journal} {Physical Chemistry Chemical Physics}\ }\textbf {\bibinfo {volume} {22}},\ \bibinfo {pages} {16760} (\bibinfo {year} {2020})}\BibitemShut {NoStop}%
\bibitem [{\citenamefont {Ozmusul}\ \emph {et~al.}(2005)\citenamefont {Ozmusul}, \citenamefont {Picu}, \citenamefont {Sternstein},\ and\ \citenamefont {Kumar}}]{ozmusul_lattice_2005}%
  \BibitemOpen
  \bibfield  {author} {\bibinfo {author} {\bibfnamefont {M.~S.}\ \bibnamefont {Ozmusul}}, \bibinfo {author} {\bibfnamefont {C.~R.}\ \bibnamefont {Picu}}, \bibinfo {author} {\bibfnamefont {S.~S.}\ \bibnamefont {Sternstein}},\ and\ \bibinfo {author} {\bibfnamefont {S.~K.}\ \bibnamefont {Kumar}},\ }\bibfield  {title} {\bibinfo {title} {Lattice {{Monte Carlo Simulations}} of {{Chain Conformations}} in {{Polymer Nanocomposites}}},\ }\href {https://doi.org/10.1021/ma0474731} {\bibfield  {journal} {\bibinfo  {journal} {Macromolecules}\ }\textbf {\bibinfo {volume} {38}},\ \bibinfo {pages} {4495} (\bibinfo {year} {2005})}\BibitemShut {NoStop}%
\bibitem [{\citenamefont {Shui}\ \emph {et~al.}(2021)\citenamefont {Shui}, \citenamefont {Huang}, \citenamefont {Wei}, \citenamefont {Chen}, \citenamefont {Song}, \citenamefont {Sun}, \citenamefont {Lu},\ and\ \citenamefont {Liu}}]{shui_intrinsic_2021}%
  \BibitemOpen
  \bibfield  {author} {\bibinfo {author} {\bibfnamefont {Y.}~\bibnamefont {Shui}}, \bibinfo {author} {\bibfnamefont {L.}~\bibnamefont {Huang}}, \bibinfo {author} {\bibfnamefont {C.}~\bibnamefont {Wei}}, \bibinfo {author} {\bibfnamefont {J.}~\bibnamefont {Chen}}, \bibinfo {author} {\bibfnamefont {L.}~\bibnamefont {Song}}, \bibinfo {author} {\bibfnamefont {G.}~\bibnamefont {Sun}}, \bibinfo {author} {\bibfnamefont {A.}~\bibnamefont {Lu}},\ and\ \bibinfo {author} {\bibfnamefont {D.}~\bibnamefont {Liu}},\ }\bibfield  {title} {\bibinfo {title} {Intrinsic properties of the matrix and interface of filler reinforced silicone rubber: {{An}} in situ {{Rheo-SANS}} and constitutive model study},\ }\href {https://doi.org/10.1016/j.coco.2020.100547} {\bibfield  {journal} {\bibinfo  {journal} {Composites Communications}\ }\textbf {\bibinfo {volume} {23}},\ \bibinfo {pages} {100547} (\bibinfo {year} {2021})}\BibitemShut {NoStop}%
\bibitem [{\citenamefont {M{\'o}cz{\'o}}\ and\ \citenamefont {Puk{\'a}nszky}(2016)}]{moczo_particulate_2016}%
  \BibitemOpen
  \bibfield  {author} {\bibinfo {author} {\bibfnamefont {J.}~\bibnamefont {M{\'o}cz{\'o}}}\ and\ \bibinfo {author} {\bibfnamefont {B.}~\bibnamefont {Puk{\'a}nszky}},\ }\bibfield  {title} {\bibinfo {title} {{Particulate Fillers in Thermoplastics}}\ }(\bibinfo  {publisher} {Springer Berlin Heidelberg},\ \bibinfo {address} {Heidelberg},\ \bibinfo {year} {2016})\ pp.\ \bibinfo {pages} {1--43}\BibitemShut {NoStop}%
\bibitem [{\citenamefont {Lin}\ \emph {et~al.}(2021{\natexlab{a}})\citenamefont {Lin}, \citenamefont {Frischknecht},\ and\ \citenamefont {Riggleman}}]{lin_chain_2021}%
  \BibitemOpen
  \bibfield  {author} {\bibinfo {author} {\bibfnamefont {E.~Y.}\ \bibnamefont {Lin}}, \bibinfo {author} {\bibfnamefont {A.~L.}\ \bibnamefont {Frischknecht}},\ and\ \bibinfo {author} {\bibfnamefont {R.~A.}\ \bibnamefont {Riggleman}},\ }\bibfield  {title} {\bibinfo {title} {Chain and {{Segmental Dynamics}} in {{Polymer}}--{{Nanoparticle Composites}} with {{High Nanoparticle Loading}}},\ }\href {https://doi.org/10.1021/acs.macromol.1c00206} {\bibfield  {journal} {\bibinfo  {journal} {Macromolecules}\ }\textbf {\bibinfo {volume} {54}},\ \bibinfo {pages} {5335} (\bibinfo {year} {2021}{\natexlab{a}})}\BibitemShut {NoStop}%
\bibitem [{\citenamefont {Afzal}\ \emph {et~al.}(2021)\citenamefont {Afzal}, \citenamefont {Browning}, \citenamefont {Goldberg}, \citenamefont {Halls}, \citenamefont {Gavartin}, \citenamefont {Morisato}, \citenamefont {Hughes}, \citenamefont {Giesen},\ and\ \citenamefont {Goose}}]{afzal_HighThroughput_2021}%
  \BibitemOpen
  \bibfield  {author} {\bibinfo {author} {\bibfnamefont {M.~A.~F.}\ \bibnamefont {Afzal}}, \bibinfo {author} {\bibfnamefont {A.~R.}\ \bibnamefont {Browning}}, \bibinfo {author} {\bibfnamefont {A.}~\bibnamefont {Goldberg}}, \bibinfo {author} {\bibfnamefont {M.~D.}\ \bibnamefont {Halls}}, \bibinfo {author} {\bibfnamefont {J.~L.}\ \bibnamefont {Gavartin}}, \bibinfo {author} {\bibfnamefont {T.}~\bibnamefont {Morisato}}, \bibinfo {author} {\bibfnamefont {T.~F.}\ \bibnamefont {Hughes}}, \bibinfo {author} {\bibfnamefont {D.~J.}\ \bibnamefont {Giesen}},\ and\ \bibinfo {author} {\bibfnamefont {J.~E.}\ \bibnamefont {Goose}},\ }\bibfield  {title} {\bibinfo {title} {High-{{Throughput Molecular Dynamics Simulations}} and {{Validation}} of {{Thermophysical Properties}} of {{Polymers}} for {{Various Applications}}},\ }\href {https://doi.org/10.1021/acsapm.0c00524} {\bibfield  {journal} {\bibinfo  {journal} {ACS Applied Polymer Materials}\ }\textbf {\bibinfo {volume} {3}},\ \bibinfo {pages} {620} (\bibinfo {year}
  {2021})}\BibitemShut {NoStop}%
\bibitem [{\citenamefont {Bowman}\ \emph {et~al.}(2019)\citenamefont {Bowman}, \citenamefont {Mun}, \citenamefont {Nouranian}, \citenamefont {Huddleston}, \citenamefont {Gwaltney}, \citenamefont {Baskes},\ and\ \citenamefont {Horstemeyer}}]{bowman_free_2019}%
  \BibitemOpen
  \bibfield  {author} {\bibinfo {author} {\bibfnamefont {A.~L.}\ \bibnamefont {Bowman}}, \bibinfo {author} {\bibfnamefont {S.}~\bibnamefont {Mun}}, \bibinfo {author} {\bibfnamefont {S.}~\bibnamefont {Nouranian}}, \bibinfo {author} {\bibfnamefont {B.~D.}\ \bibnamefont {Huddleston}}, \bibinfo {author} {\bibfnamefont {S.~R.}\ \bibnamefont {Gwaltney}}, \bibinfo {author} {\bibfnamefont {M.~I.}\ \bibnamefont {Baskes}},\ and\ \bibinfo {author} {\bibfnamefont {M.~F.}\ \bibnamefont {Horstemeyer}},\ }\bibfield  {title} {\bibinfo {title} {Free volume and internal structural evolution during creep in model amorphous polyethylene by {{Molecular Dynamics}} simulations},\ }\href {https://doi.org/10.1016/j.polymer.2019.02.060} {\bibfield  {journal} {\bibinfo  {journal} {Polymer}\ }\textbf {\bibinfo {volume} {170}},\ \bibinfo {pages} {85} (\bibinfo {year} {2019})}\BibitemShut {NoStop}%
\bibitem [{\citenamefont {Estridge}(2018)}]{estridge_effects_2018}%
  \BibitemOpen
  \bibfield  {author} {\bibinfo {author} {\bibfnamefont {C.~E.}\ \bibnamefont {Estridge}},\ }\bibfield  {title} {\bibinfo {title} {The effects of competitive primary and secondary amine reactivity on the structural evolution and properties of an epoxy thermoset resin during cure: {{A}} molecular dynamics study},\ }\href {https://doi.org/10.1016/j.polymer.2018.02.062} {\bibfield  {journal} {\bibinfo  {journal} {Polymer}\ }\textbf {\bibinfo {volume} {141}},\ \bibinfo {pages} {12} (\bibinfo {year} {2018})}\BibitemShut {NoStop}%
\bibitem [{\citenamefont {Karnes}\ \emph {et~al.}(2020)\citenamefont {Karnes}, \citenamefont {Weisgraber}, \citenamefont {Oakdale}, \citenamefont {Mettry}, \citenamefont {Shusteff},\ and\ \citenamefont {Biener}}]{karnes_network_2020}%
  \BibitemOpen
  \bibfield  {author} {\bibinfo {author} {\bibfnamefont {J.~J.}\ \bibnamefont {Karnes}}, \bibinfo {author} {\bibfnamefont {T.~H.}\ \bibnamefont {Weisgraber}}, \bibinfo {author} {\bibfnamefont {J.~S.}\ \bibnamefont {Oakdale}}, \bibinfo {author} {\bibfnamefont {M.}~\bibnamefont {Mettry}}, \bibinfo {author} {\bibfnamefont {M.}~\bibnamefont {Shusteff}},\ and\ \bibinfo {author} {\bibfnamefont {J.}~\bibnamefont {Biener}},\ }\bibfield  {title} {\bibinfo {title} {On the {{Network Topology}} of {{Cross-Linked Acrylate Photopolymers}}: {{A Molecular Dynamics Case Study}}},\ }\href {https://doi.org/10.1021/acs.jpcb.0c05319} {\bibfield  {journal} {\bibinfo  {journal} {The Journal of Physical Chemistry B}\ }\textbf {\bibinfo {volume} {124}},\ \bibinfo {pages} {9204} (\bibinfo {year} {2020})}\BibitemShut {NoStop}%
\bibitem [{\citenamefont {Voth}(2008)}]{voth_CoarseGraining_2008}%
  \BibitemOpen
  \bibfield  {author} {\bibinfo {author} {\bibfnamefont {G.~A.}\ \bibnamefont {Voth}},\ }\href@noop {} {\emph {\bibinfo {title} {Coarse-{{Graining}} of {{Condensed Phase}} and {{Biomolecular Systems}}}}}\ (\bibinfo  {publisher} {CRC Press},\ \bibinfo {address} {Boca Raton},\ \bibinfo {year} {2008})\BibitemShut {NoStop}%
\bibitem [{\citenamefont {Pavlov}\ and\ \citenamefont {Khalatur}(2016)}]{pavlov_fully_2016}%
  \BibitemOpen
  \bibfield  {author} {\bibinfo {author} {\bibfnamefont {A.~S.}\ \bibnamefont {Pavlov}}\ and\ \bibinfo {author} {\bibfnamefont {P.~G.}\ \bibnamefont {Khalatur}},\ }\bibfield  {title} {\bibinfo {title} {Fully atomistic molecular dynamics simulation of nanosilica-filled crosslinked polybutadiene},\ }\href {https://doi.org/10.1016/j.cplett.2016.04.061} {\bibfield  {journal} {\bibinfo  {journal} {Chemical Physics Letters}\ }\textbf {\bibinfo {volume} {653}},\ \bibinfo {pages} {90} (\bibinfo {year} {2016})}\BibitemShut {NoStop}%
\bibitem [{\citenamefont {Alessandri}\ \emph {et~al.}(2021)\citenamefont {Alessandri}, \citenamefont {Gr{\"u}newald},\ and\ \citenamefont {Marrink}}]{alessandri_martini_2021}%
  \BibitemOpen
  \bibfield  {author} {\bibinfo {author} {\bibfnamefont {R.}~\bibnamefont {Alessandri}}, \bibinfo {author} {\bibfnamefont {F.}~\bibnamefont {Gr{\"u}newald}},\ and\ \bibinfo {author} {\bibfnamefont {S.~J.}\ \bibnamefont {Marrink}},\ }\bibfield  {title} {\bibinfo {title} {The {{Martini Model}} in {{Materials Science}}},\ }\href {https://doi.org/10.1002/adma.202008635} {\bibfield  {journal} {\bibinfo  {journal} {Advanced Materials}\ }\textbf {\bibinfo {volume} {33}},\ \bibinfo {pages} {2008635} (\bibinfo {year} {2021})}\BibitemShut {NoStop}%
\bibitem [{\citenamefont {Karnes}\ \emph {et~al.}(2023)\citenamefont {Karnes}, \citenamefont {Weisgraber}, \citenamefont {Cook}, \citenamefont {Wang}, \citenamefont {Crowhurst}, \citenamefont {Fox}, \citenamefont {Harris}, \citenamefont {Oakdale}, \citenamefont {Faller},\ and\ \citenamefont {Shusteff}}]{karnes_isolating_2023}%
  \BibitemOpen
  \bibfield  {author} {\bibinfo {author} {\bibfnamefont {J.~J.}\ \bibnamefont {Karnes}}, \bibinfo {author} {\bibfnamefont {T.~H.}\ \bibnamefont {Weisgraber}}, \bibinfo {author} {\bibfnamefont {C.~C.}\ \bibnamefont {Cook}}, \bibinfo {author} {\bibfnamefont {D.~N.}\ \bibnamefont {Wang}}, \bibinfo {author} {\bibfnamefont {J.~C.}\ \bibnamefont {Crowhurst}}, \bibinfo {author} {\bibfnamefont {C.~A.}\ \bibnamefont {Fox}}, \bibinfo {author} {\bibfnamefont {B.~S.}\ \bibnamefont {Harris}}, \bibinfo {author} {\bibfnamefont {J.~S.}\ \bibnamefont {Oakdale}}, \bibinfo {author} {\bibfnamefont {R.}~\bibnamefont {Faller}},\ and\ \bibinfo {author} {\bibfnamefont {M.}~\bibnamefont {Shusteff}},\ }\bibfield  {title} {\bibinfo {title} {Isolating {{Chemical Reaction Mechanism}} as a {{Variable}} with {{Reactive Coarse-Grained Molecular Dynamics}}: {{Step-Growth}} versus {{Chain-Growth Polymerization}}},\ }\bibfield  {journal} {\bibinfo  {journal} {Macromolecules}\ }\href {https://doi.org/10.1021/acs.macromol.2c02069}
  {10.1021/acs.macromol.2c02069} (\bibinfo {year} {2023})\BibitemShut {NoStop}%
\bibitem [{\citenamefont {Kremer}\ and\ \citenamefont {Grest}(1990)}]{kremer_dynamics_1990}%
  \BibitemOpen
  \bibfield  {author} {\bibinfo {author} {\bibfnamefont {K.}~\bibnamefont {Kremer}}\ and\ \bibinfo {author} {\bibfnamefont {G.~S.}\ \bibnamefont {Grest}},\ }\bibfield  {title} {\bibinfo {title} {Dynamics of entangled linear polymer melts:\, {{A}} molecular-dynamics simulation},\ }\href {https://doi.org/10.1063/1.458541} {\bibfield  {journal} {\bibinfo  {journal} {The Journal of Chemical Physics}\ }\textbf {\bibinfo {volume} {92}},\ \bibinfo {pages} {5057} (\bibinfo {year} {1990})}\BibitemShut {NoStop}%
\bibitem [{\citenamefont {Lin}\ \emph {et~al.}(2021{\natexlab{b}})\citenamefont {Lin}, \citenamefont {Frischknecht}, \citenamefont {Winey},\ and\ \citenamefont {Riggleman}}]{lin_effect_2021}%
  \BibitemOpen
  \bibfield  {author} {\bibinfo {author} {\bibfnamefont {E.~Y.}\ \bibnamefont {Lin}}, \bibinfo {author} {\bibfnamefont {A.~L.}\ \bibnamefont {Frischknecht}}, \bibinfo {author} {\bibfnamefont {K.~I.}\ \bibnamefont {Winey}},\ and\ \bibinfo {author} {\bibfnamefont {R.~A.}\ \bibnamefont {Riggleman}},\ }\bibfield  {title} {\bibinfo {title} {Effect of surface properties and polymer chain length on polymer adsorption in solution},\ }\href {https://doi.org/10.1063/5.0052121} {\bibfield  {journal} {\bibinfo  {journal} {The Journal of Chemical Physics}\ }\textbf {\bibinfo {volume} {155}},\ \bibinfo {pages} {034701} (\bibinfo {year} {2021}{\natexlab{b}})}\BibitemShut {NoStop}%
\bibitem [{\citenamefont {Thompson}\ \emph {et~al.}(2021)\citenamefont {Thompson}, \citenamefont {Aktulga}, \citenamefont {Berger}, \citenamefont {Bolintineanu}, \citenamefont {Michael~Brown}, \citenamefont {Crozier}, \citenamefont {{in 't Veld}}, \citenamefont {Kohlmeyer}, \citenamefont {Moore}, \citenamefont {Nguyen}, \citenamefont {Shan}, \citenamefont {Stevens}, \citenamefont {Tranchida}, \citenamefont {Trott},\ and\ \citenamefont {Plimpton}}]{thompson_lammps_2021}%
  \BibitemOpen
  \bibfield  {author} {\bibinfo {author} {\bibfnamefont {A.~P.}\ \bibnamefont {Thompson}}, \bibinfo {author} {\bibfnamefont {H.~M.}\ \bibnamefont {Aktulga}}, \bibinfo {author} {\bibfnamefont {R.}~\bibnamefont {Berger}}, \bibinfo {author} {\bibfnamefont {D.~S.}\ \bibnamefont {Bolintineanu}}, \bibinfo {author} {\bibfnamefont {W.}~\bibnamefont {Michael~Brown}}, \bibinfo {author} {\bibfnamefont {P.~S.}\ \bibnamefont {Crozier}}, \bibinfo {author} {\bibfnamefont {P.~J.}\ \bibnamefont {{in 't Veld}}}, \bibinfo {author} {\bibfnamefont {A.}~\bibnamefont {Kohlmeyer}}, \bibinfo {author} {\bibfnamefont {S.~G.}\ \bibnamefont {Moore}}, \bibinfo {author} {\bibfnamefont {T.~D.}\ \bibnamefont {Nguyen}}, \bibinfo {author} {\bibfnamefont {R.}~\bibnamefont {Shan}}, \bibinfo {author} {\bibfnamefont {M.}~\bibnamefont {Stevens}}, \bibinfo {author} {\bibfnamefont {J.}~\bibnamefont {Tranchida}}, \bibinfo {author} {\bibfnamefont {C.}~\bibnamefont {Trott}},\ and\ \bibinfo {author} {\bibfnamefont {S.~J.}\ \bibnamefont {Plimpton}},\
  }\bibfield  {title} {\bibinfo {title} {{{LAMMPS}} - {{A}} flexible simulation tool for particle-based materials modeling at the atomic, meso, and continuum scales},\ }\href {https://doi.org/10.1016/j.cpc.2021.108171} {\bibfield  {journal} {\bibinfo  {journal} {Computer Physics Communications}\ ,\ \bibinfo {pages} {108171}} (\bibinfo {year} {2021})}\BibitemShut {NoStop}%
\bibitem [{\citenamefont {Torquato}\ and\ \citenamefont {Stillinger}(2007)}]{torquato_jamming_2007}%
  \BibitemOpen
  \bibfield  {author} {\bibinfo {author} {\bibfnamefont {S.}~\bibnamefont {Torquato}}\ and\ \bibinfo {author} {\bibfnamefont {F.~H.}\ \bibnamefont {Stillinger}},\ }\bibfield  {title} {\bibinfo {title} {Toward the jamming threshold of sphere packings: {{Tunneled}} crystals},\ }\href {https://doi.org/10.1063/1.2802184} {\bibfield  {journal} {\bibinfo  {journal} {Journal of Applied Physics}\ }\textbf {\bibinfo {volume} {102}},\ \bibinfo {pages} {093511} (\bibinfo {year} {2007})}\BibitemShut {NoStop}%
\bibitem [{\citenamefont {Hales}(2002)}]{hales_overview_2002}%
  \BibitemOpen
  \bibfield  {author} {\bibinfo {author} {\bibfnamefont {T.~C.}\ \bibnamefont {Hales}},\ }\href {https://doi.org/10.48550/arXiv.math/9811071} {\bibinfo {title} {An overview of the {{Kepler}} conjecture}} (\bibinfo {year} {2002}),\ \Eprint {https://arxiv.org/abs/math/9811071} {arXiv:math/9811071} \BibitemShut {NoStop}%
\bibitem [{\citenamefont {Jain}\ and\ \citenamefont {{de Pablo}}(2004)}]{jain_investigation_2004}%
  \BibitemOpen
  \bibfield  {author} {\bibinfo {author} {\bibfnamefont {T.~S.}\ \bibnamefont {Jain}}\ and\ \bibinfo {author} {\bibfnamefont {J.~J.}\ \bibnamefont {{de Pablo}}},\ }\bibfield  {title} {\bibinfo {title} {Investigation of {{Transition States}} in {{Bulk}} and {{Freestanding Film Polymer Glasses}}},\ }\bibfield  {journal} {\bibinfo  {journal} {Physical Review Letters}\ }\textbf {\bibinfo {volume} {92}},\ \href {https://doi.org/10.1103/PhysRevLett.92.155505} {10.1103/PhysRevLett.92.155505} (\bibinfo {year} {2004})\BibitemShut {NoStop}%
\bibitem [{\citenamefont {Ford}(2022)}]{ford_molecular_2022}%
  \BibitemOpen
  \bibfield  {author} {\bibinfo {author} {\bibfnamefont {A.}~\bibnamefont {Ford}},\ }\emph {\bibinfo {title} {{Molecular Dynamics Modeling of Heterogeneous Structure and Rheology of Human Lung Mucus}}},\ \href@noop {} {Ph.D. thesis},\ \bibinfo  {school} {University of North Carolina at Chapel Hill}, \bibinfo {address} {Department of Mathematics} (\bibinfo {year} {2022})\BibitemShut {NoStop}%
\bibitem [{\citenamefont {Sarkisov}\ \emph {et~al.}(2020)\citenamefont {Sarkisov}, \citenamefont {{Bueno-Perez}}, \citenamefont {Sutharson},\ and\ \citenamefont {{Fairen-Jimenez}}}]{sarkisov_materials_2020}%
  \BibitemOpen
  \bibfield  {author} {\bibinfo {author} {\bibfnamefont {L.}~\bibnamefont {Sarkisov}}, \bibinfo {author} {\bibfnamefont {R.}~\bibnamefont {{Bueno-Perez}}}, \bibinfo {author} {\bibfnamefont {M.}~\bibnamefont {Sutharson}},\ and\ \bibinfo {author} {\bibfnamefont {D.}~\bibnamefont {{Fairen-Jimenez}}},\ }\bibfield  {title} {\bibinfo {title} {Materials {{Informatics}} with {{PoreBlazer}} v4.0 and the {{CSD MOF Database}}},\ }\href {https://doi.org/10.1021/acs.chemmater.0c03575} {\bibfield  {journal} {\bibinfo  {journal} {Chemistry of Materials}\ }\textbf {\bibinfo {volume} {32}},\ \bibinfo {pages} {9849} (\bibinfo {year} {2020})}\BibitemShut {NoStop}%
\end{thebibliography}%
\end{document}



\title{Supplementary Material for ``On void formation during the simulated tensile testing of polymer-filler particle composites''}

\author{John J. Karnes}
 \email{karnes@llnl.gov}
\author{Supun S. Mohottalalage}%
\author{Amitesh Maiti}
\author{Andrew P. Saab}
\author{Todd H. Weisgraber}

\affiliation{%
 Lawrence Livermore National Laboratory\\
 5000 East Ave., Livermore, CA, 94550, USA
}%




\date{\today}



\maketitle

\section*{\label{sec:intro}Calculating voidspace distribution functions}

\begin{figure}[h!]
\includegraphics[width=3.375in]{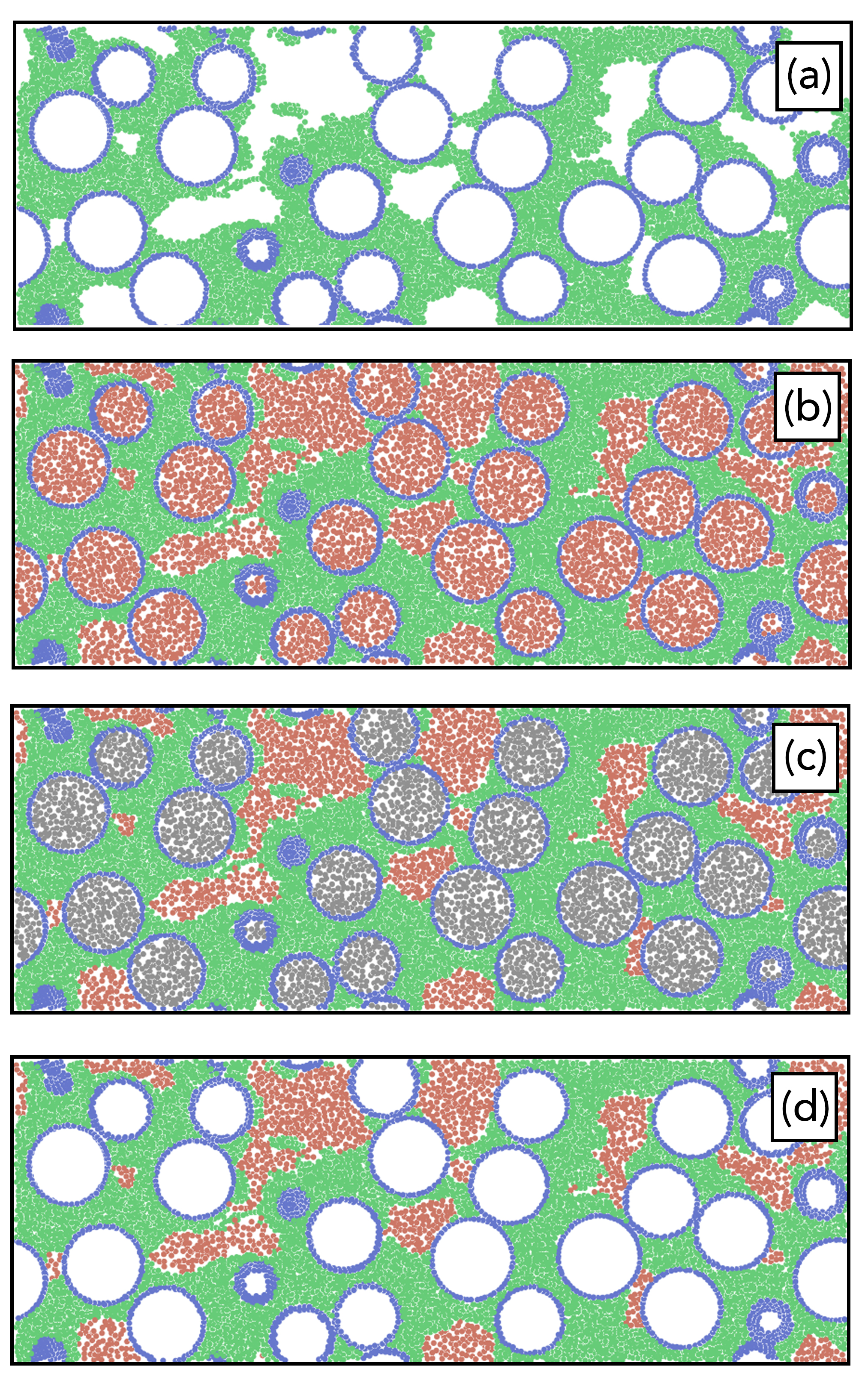}
\caption{Visual outline of the snapshot preparation for $g_v(r)$ calculation. (a) Snapshot from simulated uniaxial extension (b) insertion of dummy atoms (c) mutation of dummy atoms within FP (d) mutated dummy atoms made invisible for visual clarity.}
\label{fig:gvr}
\end{figure}

In the main text we adapt the well-known pair distribution function, $g(r)$, to enable spatial quantification of voids or delaminations that form during the simulated tensile testing of polymer composites. In this Supplementary Material we outline the modification of simulation snapshots to enable the calculation of the voidspace distribution function $g_v(r)$. This overview refers to operations within the LAMMPS environment~\cite{plimpton_fast_1995,thompson_lammps_2021} and we refer the interested reader to LAMMPS documentation for additional details regarding the individual commands mentioned. Figure \ref{fig:gvr} is a visual outline of the steps described below using a 2-D slice of an elongated simulation cell for clarity.  

\subsection{\label{sec:model} Filling the voids}

Our approach involves filling the appropriate void spaces with a unique particle type (`dummy atoms') that will permit calculation of $g_v(r)$. We first read in a simulation snapshot and allow for two new atom types, 
\begin{verbatim}
read_data dataOut.data extra/atom/types 2
\end{verbatim} 

and insert new dummy atoms by looping over a command such as

\begin{verbatim}
create_atoms 5 random 10000 121717 //
NULL overlap 1.05 maxtry 10000
\end{verbatim}

This action will fill all of the simulation voidspaces with dummy atoms of type 5 as in Figure \ref{fig:gvr}(b). However, since we represent the filler particles (FP) as hollow shells (i.e. Fibonacci spheres), the insides of the FP will also be filled with dummy atoms. 

\subsection{\label{sec:model} Clean out the filler particles}

Dummy atoms within the filler particles will make calculation of $g_v(r)$ difficult since not only will the reference atom at the center of each FP be surrounded by dummy atoms, the reference atom is also likely to `see' dummy particles within nearby FP. Adjusting the range of $g_v(r)$ is not sufficient. We use the LAMMPS command \verb|fix bond/react|~\cite{gissinger_modeling_2017,gissinger_reacter_2020} to mutate dummy atoms inside the FP into an atom type that will be ignore by the calculation of $g_v(r)$, in this example \verb|type 6|. 

\begin{verbatim}
fix fbr all bond/react stabilization yes // 
nvt_grp 0.01 react myrxn cleanup //
1 1.0 ${FP_rad} mol1 mol2 map.txt //
stabilize_steps 0
\end{verbatim}

\subsection{\label{sec:model} Compute $g_v(r)$}

The modified snapshot (Figure \ref{fig:gvr}(c)) now permits calculation of conventional $g(r)$, such as the proximity of FP to other FP or spatial quantification of voidspace by referencing the dummy atoms inserted by the preceding approach. For example,

\begin{verbatim}
compute rdfs all rdf 100 1 1 3*4 5 //
cutoff 19.5
fix prdf all ave/time 1 1 1 //
c_rdfs[*] file rdf.txt mode vector
\end{verbatim}

The above command calculates several $g(r)$, including $g_{\text{FP-FP}}(r)$, $g_{\text{FP-polymer}}(r)$, and $g_v(r)$ with the centers of the FP used as reference atoms.

%